\documentstyle[epsf,sprocl,hyperref]{article} 
\begin{document}
\begin{flushright}
{hep-th/9809103}

AEI-093
\bigskip
\end{flushright}
\title{SUPERMEMBRANES AND M(ATRIX) THEORY\footnote{Lectures given
by H.~Nicolai at the Trieste Spring School on Non-Perturbative
Aspects of String Theory and Supersymmetric Gauge Theories, 
23 - 31 March 1998.}}

\author{Hermann Nicolai and Robert Helling}
\address{ Max Planck Institut f\"ur Gravitationsphysik\\
          Albert-Einstein-Institut\\ 
          Schlaatzweg 1\\
          D-14473 Potsdam, Germany}

\maketitle

\begin{abstract}
In these lectures, we review the $d=11$ supermembrane and 
supersymmetric matrix models at an introductory level.
We also discuss some more recent developments in connection
with non-perturbative string theory. 
\end{abstract}

% \magnification = \magstep1
% Including memstyle.tex
% Including ../robsenv.tex
\bibliographystyle{hunsrt}
\parindent=0cm

\font\sans=cmss10

\font\gross=cmr10 scaled \magstep1

\font\riesig=cmr10 scaled \magstep2

\font\geom=cmssbx10

\font\mengen=bbm10
%\font\csc=cmcsc10
\font\ninerm=cmr9
%\font\ninebf=cmr9 % We want references in rm!
\font\ninebf=cmb9
%\font\fraktur=cmfrak

\edef\ignore#1{}

\def\Det{\hbox{Det}}
\def\Tr{\hbox{Tr}}
\def\tr{\hbox{tr}}
\def\part#1#2{{\partial #1\over\partial #2}}

\def\br#1#2{{#1\over #2}}

\def\frac#1#2{{#1\over #2}}

\def\caret{{\tt\char'136}}
\def\nach{\rightarrow}

\def\sla#1{\hbox to 0cm{\kern 0.3ex/\hss}#1}

\def\MN{\hbox{\mengen N}}
\def\MZ{\hbox{\mengen Z}}
\def\MR{\hbox{\mengen R}}
\def\MC{\hbox{\mengen C}}
\def\ID{\hbox{\mengen 1}}
\def\Tr{\mathop{\hbox{Tr}}}
\def\kuller{$\bullet$}
\def\quer{\bar}
\newcount\figcount \figcount=0
\def\fig#1#2{\global\advance\figcount by1\midinsert\vskip #2
\centerline{Fig.\ \the\figcount :#1}\endinsert}
\def\psfig#1#2{\global\advance\figcount
by 1 \begin{figure}[t]\vbox{\centerline{\epsfbox{#2}}
\centerline{Fig.\ \the\figcount :#1}}\end{figure}}
\def\psfigs#1#2#3{\global\advance\figcount
by 1 \begin{figure}[t]\vbox{\centerline{\epsfbox{#2}\epsfbox{#3}}
\centerline{Fig.\ \the\figcount :#1}}\end{figure}}
\def\insertsomestuff#1#2{\global\advance\figcount
by 1 \begin{figure}[t]\vbox{{{#2}}
\par
\centerline{Fig.\ \the\figcount :#1}}\end{figure}}
\newcount\chapterno
\newcount\glno
\def\neueseitevorchapter{\vfill\supereject} % Can be set to \par by \draft
\def\chapter#1{\global\advance\chapterno by 1\glno =0\section{#1}}
\def\gln#1{\global\advance\glno by 1\xdef#1{(\the\chapterno
.\the\glno)}\eqno {(\the\chapterno .\the\glno)}}
\def\egln#1{\global\advance\glno by 1\xdef#1{(\the\chapterno
.\the\glno)}& {(\the\chapterno .\the\glno)}}

% Hier stehen ein paar plain Macros, die LaTeX nicht kennt.
\newif\dtap

\def\eqalign#1{\null\,\vcenter{\openup1\jot \mathsurround=0pt
\everycr={}\tabskip=0pt \halign{\strut\hfil$\displaystyle{##}$&$
\displaystyle{{}##}$\hfil \crcr#1\crcr}}\,}

\def\olign{\tabskip\z@skip\everycr{}} % restore inside \displ@y

\def\displ{\global\dtaptrue \openup1\jot \mathsurrount=0pt
\everycr{\noalign{\ifdtap \global\dtapfalse\vskip-\lineskiplimit
\vskip\normallineskiplimit \else \penalty\interdisplaylinepenalty\fi}}}

\def\eqalignno#1{\displ \tabskip=\centering \halign
to\displaywidth{\hfil\olign\displaystyle{##} $\tabskip=0pt
&$\olign\displaystyle{{}##}$\hfil\tabskip=\centering
&\llap{$\olign##$}\tabskip =0pt\crcr#1\crcr}}

\let\epsffile=\epsfbox
\newwrite\auxfile\openout\auxfile=lectures.preaux
\newcount\referenz
\jot=10pt  % Make things more LaTeXy
\def\draft{\def\vielluft{\bigskip}\def\neueseitevorchapter{\par}}
\newif\ifcommentsindvi
\newif\ifcommentsinlog
\def\comment#1{{\ifcommentsindvi\ninebf[#1]\fi\ifcommentsinlog\message{Author
comment:#1}\fi}}
\def\z#1{\zeta^{#1}}
\def\G#1{\Gamma^{#1}}
\def\X#1{X^{#1}}
\def\O{\theta}
\def\E#1#2{E_{#1}{}^{#2}}
\def\d#1{\partial_{#1}}
\def\f#1#2#3{f_{#1#2}{}^{#3}}

\def\emph#1{{\bf #1}}
\def\breakhere{}

\hyphenation{su-per-mem-brane}

\commentsindvitrue   %true durch false ersetzen, wenn die Kommentare
                     %nicht in den Ausdruck sollen.
\commentsinlogtrue   %true durch false ersetzen, wenn die Kommentare
                     %nicht in die LaTeX-Ausgabe sollen ('silent run')
\def\eps{.k.eps}

%
%HIER BEGINNT DER RICHTIGE TEXT
%
\chapter{Introduction}
The purpose of these lectures is to give an introduction to
supermembranes, with special emphasis on the maximally extended $d=11$
theory, to supersymmetric matrix models, and to explain the relation between
these theories. In doing so, we will not only review ``old'' results,
but also discuss some of the more recent developments. Although we
cannot give all the technical details, we will try to be pedagogical 
and to concentrate on what we consider the salient and most important points.

As is by now well known, the large $N$ limit of the maximally supersymmetric 
$SU(N)$ matrix model of supersymmetric quantum mechanics is a serious
candidate for M-Theory, the still elusive theory unifying $d=11$
supergravity and superstring theory at the non-perturbative level.
The very same model had been encountered more than ten years ago in a
study of supermembranes in the light-cone gauge, where the
supersymmetric $SU(N)$ matrix model was proposed as a non-perturbative 
regularization of the supermembrane, from which the
full quantum supermembrane can be obtained by taking the limit
$N\rightarrow\infty$. Notwithstanding some subtle interpretational
distinctions, there is thus no real difference between what is now
called M(atrix) theory and the quantum supermembrane, provided a
unique large $N$ limit can be shown to exist. Rather, the remarkable
fact is that the same model can be arrived at in two so different ways. 
A crucial new insight occasioned by advances in D-branes, which will 
receive due emphasis in these lectures, is that the quantum supermembrane 
is a second quantized theory from the very outset.

\insertsomestuff{Two approaches to M(atrix) Theory}{
$$\vbox{\halign{\strut\hfill#&#\hfill\cr
\hidewidth\gross Supermembranes\span\hidewidth\cr
{}&$\Bigg \updownarrow$\hbox to 0pt{\ninerm (1988)\hss}\cr
\hidewidth\gross M(atrix)-Theory\span\hidewidth\cr
{}&$\Bigg \updownarrow$\hbox to 0pt{\ninerm (1996)\hss}\cr
\hidewidth\gross Mechanics of D0-particles\span\hidewidth\cr
}}$$
}

Despite the recent excitement, however, we do not think that
M(atrix) theory and the $d=11$ supermembrane in their present
incarnation are already the final answer in the search for M-Theory,
even though they probably are important pieces of the puzzle. There
are still too many ingredients missing that we would expect the final 
theory to possess. For one thing, we would expect a true theory of
quantum gravity to exhibit certain pregeometrical features
corresponding to a ``dissolution'' of space-time and the emergence
of some kind of non-commutative geometry at short distances;
although the matrix model does achieve that to some extent 
by replacing commuting coordinates by non-commuting matrices,
it seems to us that a still more radical departure from
conventional ideas about space and time may be required in order
to arrive at a truly background independent formulation (the matrix 
model ``lives'' in nine {\em flat} transverse dimensions only). 
Furthermore, there should exist some huge and so far completely hidden 
symmetries generalizing not only the duality symmetries of extended 
supergravity and string theory, but also the principles 
underlying general relativity.

\chapter{Basics of supermembranes}
Since there exist several reviews of supermembrane theory
\cite{Bergshoeff:1988qx,Duff:1990sc,Duff:1996zn,Stelle:1996tz,deWit:1997sx}
we here only summarize the basic facts, referring readers there for
full details and complementary points of view. By definition, a
(super-)$p$-brane is a $p$-dimensional extended object moving in a
target (super)space whose bosonic $d$-dimensional subspace can be
curved (subject to certain consistency conditions), but will be taken
to be flat $\MR^d$ for simplicity here. The dimension of the fermionic
subspace is quite generally determined by the number of components of
a spinor in $d$ dimensions (possibly with extra factors of 1/2 or 1/4
for Majorana and/or Weyl spinors). Unlike the bosonic case, where $p$
and $d$ can be chosen more or less at will, the number of
possibilities for supersymmetric extended objects is quite limited
(the allowed dimensions are listed in a ``brane scan''
\cite{Achucarro:1987nc}). This is essentially because the number of
spinor components grows exponentially with dimension, whereas the
number of components of a vector grows only linearly, and it becomes
impossible to match bosonic and fermionic degrees of freedom once $d$
gets too large.

We parametrize the $(p+1)$-dimensional world volume by local coordinates 
$$
\z i=(\z 0,\z r)\equiv (\tau,\sigma^r)
$$
with indices $r, s,\ldots = 1,\ldots ,p$ labeling the spacelike
coordinates on the $p$-brane. Accordingly, each point in the world 
volume is mapped to a point in target superspace according to
$$
\zeta \longmapsto (\X\mu (\zeta), \O_\alpha (\zeta))
$$
The space time indices $\mu, \nu,\ldots$ run over $0,1,\ldots, d-1$,
and the indices $\alpha, \beta$ label the components of a spinor in
$d$ dimensions.  
In these
lectures we will almost exclusively be concerned with $p=2$,
i.e. supermem\-branes, which, even as classical theories, can only
exist in target spaces of dimension 4,5,7 and 11. Quantum
mechanically, there will be further restrictions, just as for the
superstring, such that $d=11$ is presumably the only viable candidate 
for a consistent quantum supermembrane. Therefore $d=11$ is the most 
interesting case, also because it is related to the unique and maximal 
supergravity theory in eleven dimensions; \cite{Cremmer:1978km} in this
case, there are 32 real fermionic coordinates $\O_\alpha$
(i.e. $\alpha,\beta,\ldots = 1,\ldots, 32$) corresponding to the
components of a Majorana spinor in eleven dimensions. Note, however,
that from the world volume point of view, $\theta_\alpha$ transforms
as a scalar. This is a general feature of Green Schwarz type actions.

To construct the action of the supermembrane, one starts from the 
well-known Nambu-Goto action principle:
$$\hbox{action} = \hbox{world volume}$$
We thus first of all need an expression for the induced metric on
the world volume allowing us to define a volume element.
This is simply obtained by pulling back the target space
metric using the coordinate functions:
$$
g_{ij}(X,\O) = \E i\mu\E j\nu \eta_{\mu\nu}
$$
where the $\{ \E i\mu | i=0,1,2 \}$ form a dreibein 
tangent to the world volume. For a target {\em super}space, this
dreibein becomes a supervielbein with extra fermionic
components $\d i\theta$. The bosonic part of the vielbein
for the supermembrane reads
$$
\E i\mu := \d i\X\mu + \bar\O\G\mu\d i\O.
$$
The $32 \times 32$ matrices $\Gamma^\mu$ generate the target 
space Clifford algebra, i.e.
$$
\{ \Gamma^\mu , \Gamma^\nu \} = 2\eta^{\mu\nu}
$$
These ingredients are all that is needed to write down 
the supermembrane action in a flat eleven-dimensional target space
\cite{Bergshoeff:1988ui,Bergshoeff:1988qx}  
$$\eqalign{
{\cal L} &= -\sqrt{-g(X,\O)} - \cr
&\qquad -\epsilon^{ijk}\biggl[\br 12 \d i\X\mu
(\d j \X\nu +\bar\O\G\nu\d j\O )+\br 16\bar\O\G\mu\d i\O\bar\O\G\nu\d
j\O\biggr]\bar\O\Gamma_{\mu\nu}\d k\O\cr   }\gln\lagrangian
$$ 
which represents a generalization of the Green Schwarz action for the 
superstring (which exists for target space dimensions $d=3,4,6$ and 10). 
This action is not so easy to guess, and is most conveniently
derived in the general superspace formulation of \cite{Bergshoeff:1988ui}
which also exists for non-trivial backgrounds. The second term can 
be interpreted as a WZW term in the target superspace.\cite{Henneaux:1985mh} 

Of course, this action will have extra terms once a non-trivial 
supergravity background is switched on; consistency then requires 
that the background fields satisfy the equations of motion of 
$d=11$ supergravity \cite{Bergshoeff:1988ui,Bergshoeff:1988qx}
(see \cite{deWit:1998tk} for a more recent analysis of the
supermembrane in a curved background). 
Note also that for simplicity we have set the inverse
membrane tension $T_p$, which multiplies the action to unity.
This parameter can be easily put back into all
formulas by simple dimensional arguments. Let us mention 
that we can alternatively treat the world volume metric $g_{ij}$ 
as an independent variable on the $(p+1)$ dimensional world volume. 
Solving the equations of motion that follow from
$$
{\cal L} = -\br 12\sqrt{-g} g^{ij}\E i\mu E_{j\mu}+\br 12
(p-1)\sqrt{-g}
$$ 
for the world volume metric, and substituting the on-shell metric,
brings us back to the previous Nambu-Goto-like action. Unfortunately,
this is not enough to set up a formulation of the supermembrane
which would be the world volume analog of the Ramond and Neveu-Schwarz 
formulations of the superstring.

We also record the Euler-Lagrange equations of motion following 
from the above Lagrangian
$$\eqalign{
\d i(\sqrt{-g}g^{ij}\E j\mu) &=\epsilon^{ijk}\E j\nu\d
j\bar\O\G\mu{}_\nu\d k\O,\cr
 (\ID +\Gamma)g^{ij}\sla{E_i}\d j\O&=0\cr}
$$
where the field dependent matrix
$$
\Gamma := {\epsilon^{ijk}\over 6\sqrt{-g}} \E i\mu\E j\nu\E k\rho
\Gamma_{\mu\nu\rho}
$$ 
obeys $\Gamma^2 = \ID$, whence $\ID \pm \Gamma$ are
projection operators.

The most important feature of the above action are its symmetries.
The global ones are determined by the (bosonic and fermionic)
Killing symmetries of the $d=11$ background geometry in which the 
supermembrane moves. For a {\em flat} $d=11$ target space, they correspond 
to the target space super-Poincar\'e transformations 
$$
\eqalign{\delta\X\mu &= a^\mu
+\omega^{\mu\nu}X_\nu-\bar\epsilon\G\mu\O\cr
\delta\O &= \br 14 \omega_{\mu\nu}\G{\mu\nu}\O+\epsilon\cr}\gln\globalsusy
$$
As is well known, the global invariances give rise to
conserved charges. For instance, the current associated with
$d=11$ target space supersymmetry reads
$$\eqalign{J^i =& -2 \sqrt{-g}g^{ij}\sla {E_j}\O-\epsilon^{ijk} \Bigl\{\E j\mu
\E k\nu\Gamma_{\mu\nu}\O+ \cr
&\quad+\br 43\left[\G\nu\O(\bar \O\Gamma_{\mu\nu}\d
j\O) +\Gamma_{\mu\nu}\O(\bar\O\G\nu\d j\O)\right]
(\E k\mu -\br
25\bar\O\G\mu\d k\O)\Bigr\}\cr}\gln\current$$
such that the global supersymmetry variations are generated by
the Noether supercharges
$$
Q=\int d^2\sigma J^0
$$
The other currents are given by standard expressions and are not
presented here.

The local (gauge) symmetries are associated with the invariance under
re\-pa\-ra\-met\-ri\-zat\-ions of the world volume coordinates along a vector 
field $\xi$ and a fermionic $\kappa$-symmetry
\cite{Bergshoeff:1988ui,Siegel:1983hh,Siegel:1985ys} 
$$
\eqalign{\delta\X\mu &= \xi^i\d i\X\mu +\bar\kappa (\ID-\Gamma)\G\mu
\O\cr
\delta\O &=\xi^i\d i\O+(\ID-\Gamma)\kappa\cr} 
$$
Here $\kappa$ is again a 32 component Majorana spinor. The WZW term 
in the Lagrangian is essential for $\kappa$-symmetry to hold.
As already pointed out above, the combination $(\ID-\Gamma)$ is a projector:
It eliminates half of the spinor components of $\kappa$. This
means that $\kappa$-symmetry halves the number of physical spinor
components. The $\kappa$ invariance requires the Fierz identity
$$\bar\psi_{[1}\G\mu\psi_2\bar\psi_3\Gamma_{\mu\nu}\psi_{4]}=0\gln\fierz$$
which holds only in $d=4,5,7,11$ dimensions (this identity is
analogous to the one in supersymmetric gauge theories
\cite{Green:1987sp} and the standard Green Schwarz action, which leads
to the values $d=3,4,6,10$).  This is the technical reason why
supermembranes exist only in dimensions $d=4,5,7,$ and 11. As we
already said, in these lectures we shall restrict ourselves mostly
to the case $d=11$, since this is the one relevant to M-theory.

The local invariances allow us to eliminate unphysical (i.e. off-shell) 
degrees of freedom such that on-shell, we are left with an equal number 
of bosonic and fermionic degrees of freedom, as required by supersymmetry. 
Namely, the reparametrization invariance can be used to eliminate
(on the mass-shell) three polarizations of $\X\mu$ tangential to the
world volume. This gauge is not 
the same as the light-cone gauge that we will discuss later, but
gives the right count more quickly. For the fermions, we invoke
$\kappa$ symmetry to get rid of one half of them as explained above.
Altogether, we then end up with the following count of degrees 
of freedom in the allowed dimensions for the supermembranes.
{\def\quad{\,\,\,}
$$\vbox{\halign{\vrule\strut\hfil\quad$#$\hfil\quad&\vrule\hfil\quad$#$&${}#$
\quad\hfil&\hfil\quad 
$#$&${}#$\quad\hfil&\hfil\quad$#$&${}#$\quad\hfil&\hfil\quad$#$&${}#$
\quad\hfil\vrule\cr
\noalign{\hrule} 
&d&=4&d&=5&d&=7&d&=11\cr
\noalign{\hrule}
\X\mu&4&\to 1&5&\to2&7&\to4&11&\to 8\cr
\O&4&\to 2&8&\to 4&16&\to 8&32&\to 16\cr
\noalign{\hrule}
}}$$    %Hier war noch eine hrule, aber entweder ganze box rum oder
        %gar nix.
}
%&N&=4&N&=8&N&=16&N&=32\cr}}$$
This achieves the matching on the mass shell, because, canonically,
half of the (real) spinor components must be treated as coordinates 
and the other half as momenta. Thus we get another factor of 1/2.

We had to fix all gauge degrees of freedom to make this counting work
and to exhibit the supersymmetry of the spectrum explicitly. It is 
one of the outstanding problems to find auxiliary fields which would 
also match the degrees of freedom off-shell. Solving it is probably
 as hard as finding an off-shell formulation for $N=8$ supergravity 
(still unknown after 20 years). In fact, it is likely that no 
off-shell formulation in the conventional sense (with finitely many 
auxiliary fields) exists at all. Nevertheless, some equivalent of 
an off-shell version of the supermembrane or an equally
powerful tool is needed if we want to settle the problem of
renormalizability vs. non-renormalizability once and for all,
for instance by simply excluding the existence of supersymmetric 
counterterms depending on the world volume curvature.

\chapter{Light-cone gauge}
The gauge which is traditionally chosen to analyze the physical
content of a gauge theory is the so-called lightcone gauge. The
Hamiltonian formulation of the bosonic membrane in this gauge
was given in \cite{Goldstone,Hoppe}. For the supermembrane this gauge 
was first studied in \cite{Bergshoeff:1988in,deWit:1988ig}.
To implement it, we make partial use of the world volume 
reparametrization invariance to impose the lightcone gauge condition
$$
\X +(\zeta)=X_0^++\tau\iff \d i\X += \delta_{i0},
$$
Here we have introduced the standard lightcone coordinates
$$
\X\pm =\br 1{\sqrt 2}(X^{10}\pm\X 0).
$$
We will denote transverse coordinates by $\vec X(\zeta)=\X a(\zeta)$, with
$a=1,\ldots ,9$. This has reduced the
number of bosonic coordinates from 11 to 9. Similarly, we define
$$
\Gamma^\pm = \br 1{\sqrt 2}(\G{10}\pm\G{0})
$$
Next, we use $\kappa$-symmetry to eliminate 16 of the 32 fermionic
coordinates by imposing
$$\Gamma^+\O=0.$$
With these substitutions, we find
$$\eqalign{g_{rs}&\equiv\bar g_{rs}=\d r\X a\d s\X a\equiv \d r\vec
X\cdot\d s\vec X\cr
g_{0r}&\equiv u_r=\d r\X -+\d 0\vec X\d r\vec X+\bar \O \Gamma^-\d
r\O\cr
g_{00}&= 2\d 0\X -+(\d 0\vec X)^2+2\bar\O\Gamma^-\d 0\O.\cr}$$
Now, the Lagrangian simplifies significantly:
$$
{\cal L} = -\sqrt{\bar g\Delta} +\epsilon^{rs}\d r\X
a\bar\O\Gamma^-\Gamma_a\d s\O
$$
Here, we defined $\bar g\equiv\det\bar g_{rs}$ and $\Delta\equiv
-g_{00}+u_r\bar g^{rs}u_s$. 

Since here our main interest is in studying the relation between
the supermembrane and the matrix model, we pass on to the  
Hamiltonian formulation of the supermembrane without further ado.
As a first step, let us work out the canonical momenta:
$$\eqalign{\vec P &={\partial{\cal L}\over\partial(\d 0\vec X)}
=\sqrt{\bar g\over\Delta}(\d 0\vec X-u_r\bar g^{rs}\d s\vec X)\cr
P^+&= {\partial{\cal L}\over\partial (\d 0\X -)}=\sqrt{\bar
g\over\Delta}\cr
S&= {\partial{\cal L}\over\partial (\d 0\bar\O)}=-\sqrt{\bar
g\over\Delta}\Gamma^-\O\cr}$$
After some algebra, the Hamiltonian density is found to be
$$
\eqalign{{\cal H}&\equiv\vec P\cdot \d 0\vec X+P^+\d 0\X -+\bar S\d
0\O-{\cal L}\cr
&= {\vec P^2+\bar g\over 2P^+}-\epsilon^{rs}\d r\X
a\bar\O\Gamma^-\Gamma_a\d s\O\cr}\gln\lcham
$$
The Hamiltonian is then the integral of this density over the membrane, viz.
$$
H \equiv -P^-_0 = \int_{\cal M} d^2\sigma {\cal H}(\sigma)\gln\hamiltonian
$$

We can further simplify the above expressions for the metric 
components by making use of the following residual invariance
of the gauge conditions under spatial diffeomorphisms. 
Putting $\xi^0 =0$, we transform $\sigma^r$ according to
$$
\sigma^r \mapsto \sigma^r + \xi^r(\tau,\sigma)
$$
in order to achieve 
$$u^r=0.\gln\lcgc$$ 
A little further thought (see e.g. \cite{deWit:1988ig}) 
then shows that the Hamiltonian equations in this gauge imply that
$$
\d 0P^+=0.
$$
\def\w{\sqrt{w(\sigma)}}
\def\wpr{\sqrt{w(\sigma')}}
Keeping in mind that $P^+(\sigma)$ transforms as a density under 
diffeomorphisms, we can easily solve this equation
$$
P^+(\sigma)=P_0^+\w
$$
where the function $\w$ is normalized as 
$$\int d^2\sigma \w =1.$$
$w(\sigma)$ should be viewed as the metric determinant on the 
membrane associated with a fiducial background metric $w_{rs}(\sigma)$
(this is a 2-by-2 spatial metric on the membrane itself, and should
not be confused with the Lorentzian metric on the membrane 
world volume). This metric is assumed to be non-singular, but 
otherwise arbitrary. Of course, we will have to make sure 
eventually that no physical quantity depends on this choice. 
Quite generally, this independence will follow from the invariance 
of the lightcone theory under area preserving diffeomorphisms 
(which by definition leave the metric density $\w$ invariant), and 
the fact that all the relevant quantities --- with the exception of
the Lorentz boost generators --- involve the metric $w_{rs}$ only 
through its determinant. The check will be rather more subtle
for the Lorentz generators which depend explicitly on the
associated Laplace-Beltrami operator. 

As in string theory, we can utilize the constraints to 
eliminate $X^-$. Namely, from \lcgc\ we read off 
$$
\d r X^-= -\d 0\vec X\cdot\d r \vec X - \bar\theta \Gamma^- \d r\O.\gln\xminus
$$
In contrast to string theory, we here get two equations (for
higher dimensional $p$-branes we would get $p$ equations) whose
compatibility is not automatic: the equation for $X^-$ can only be 
solved on a subspace of the full supermem\-brane phase space.
To ensure the compatibility, we must demand that the vector field 
has vanishing curl, which immediately yields the constraint
$$
\phi = \epsilon^{rs} (\d r\vec P\cdot \d s\vec X+\d r\bar\O
\Gamma^- \d s\O)\approx 0\gln\sconst
$$
Here, we adopt the standard notion of ``weakly zero'' ($\approx 0$).
\cite{Dirac67}
This means that we restrict the phase space to the submanifold where 
the constraints vanish but nevertheless, as functions on phase space, 
the constraints may have non-vanishing Poisson (or Dirac) brackets with other phase 
space variables. With the help of the canonical brackets
$$\eqalign{
\{ P^a (\sigma) , X^b (\sigma')\} &= \delta^{ab}\delta^{(2)}(\sigma - \sigma')
\cr 
\{\O_\alpha(\sigma),\bar\O_\beta(\sigma')\} &=
 {1\over 4\w} \Gamma^+_{\alpha\beta}\delta^{(2)}(\sigma-\sigma'),
\cr}\gln\canbrack
$$
it is now straightforward to verify that the constraint has 
vanishing canonical brackets with the Hamiltonian, and can thus 
be used to reduce the number of $X^a$ fields from nine to eight.
Let us also mention that for topologically non-trivial membranes,
there are extra consistency conditions corresponding to the
non-contractible cycles on the membrane. 

The fact that the constraint commutes with the Hamiltonian implies the
existence of a residual gauge symmetry of the lightcone
Hamiltonian. This is the invariance under area preserving
diffeomorphisms, which we will discuss in detail in section~5.

\chapter{Some Properties of the Lightcone Hamiltonian}
A lot about the qualitative features of
supermembrane theory can be learnt by studying some general 
properties of the Hamiltonian \lcham.
A first, and rather obvious, observation is that the zero modes 
$$
\eqalign{\vec X_0&\equiv\int d^2\sigma\w \vec X(\sigma)\cr
\O_0 &\equiv\int d^2\sigma\w \O(\sigma)\cr}
$$
do not appear in this Hamiltonian. Likewise, the center of mass momenta
$$
\eqalign{\vec P_0&\equiv \int d^2\sigma\vec P(\sigma)\cr
P_0^-&\equiv -\int d^2\sigma{\cal H}\cr}
$$
decouple from the non-zero mode degrees of freedom of the membrane.
By subtracting this contribution from the Hamiltonian, we arrive 
at the mass formula
%{\def\{{.}\def\}{.}
$$
{\cal M}^2 = -2P_0^+ P_0^- - \vec P_0^2  
= \int d^2\sigma\left\{
\frac{[\vec P^2]'+ \bar g}{\w} -2P_0^+ \epsilon^{rs}\d
rX^a\bar \O\Gamma^-\Gamma_a\d s\O \right\} \gln\lcham
$$
%}
by substituting the expression \hamiltonian\ for $P^-_0$ in terms of the
transverse degrees of freedom. The prime indicates that the zero modes
are to be omitted.
This formula contains all the non-trivial dynamics of the membrane, whereas
the center of mass motion is governed by the kinematics
of a free relativistic part\-icle. Similarly, the fact that the fermionic 
zero modes decouple will be used later to show that, if there exists
a massless state, it will give rise to precisely one massless
supermultiplet of $d=11$ supergravity.

The bosonic part of the 
Hamiltonian \lcham\ is of the standard form 
$$
H= {\cal M}^2 =T+V
$$
with the kinetic energy $T$. The potential energy $V$ is the 
integral of the density
$$
\bar g = \det_{r,s}(\d r\vec X\cdot\d s\vec X)= (\epsilon^{rs}\d r X^a\d
sX^b)^2
$$
The potential density vanishes if the surface degenerates, which 
happens when the $\vec X$'s only depend on one linear combination 
of the $\sigma^r$'s, i.e. when the membrane grows infinitely thin
(i.e. stringlike) spikes, \psfig{A membrane growing tubes}{schlaeuche\eps}
see Fig.\ \the\figcount. This instability has important consequences. 
We will come back to it below to show that this degeneracy must
be interpreted as manifestation of the second quantized 
nature of the quantum supermembrane.

The bosonic part of the above formula \lcham\ can be generalized 
to arbitrary $p$, with the result that the degeneracies of the
potential persist for higher $p$ (generally, the zero energy
configurations of a $p$-brane are those where the brane degenerates 
to an object of dimension less than $p$). It is only for $p=1$,
i.e. string theory, that the potential is confining when the zero
mode contribution is removed. This is also the only case where the
potential is quadratic ($\vec X'(\sigma))^2$. Then the Hamiltonian
describes a free system, and we recover the  well known result 
that the (super)string is an infinite set of (supersymmetric) 
harmonic oscillators. Thus, string theory is ``easy'' because it 
is a free theory, so that we can not only find a complete set of 
solutions to its equations of motion subject to various boundary 
conditions, but also quantize the theory straightforwardly.
By contrast, the membrane and the higher dimensional $p$-branes
are non-linear theories (with potentials $\propto \vec X^{2p}$). So, for 
instance, only a very  limited number of special solutions to 
the classical membrane equations of motion are known.

A related observation is that, at a purely kinematical level, 
the type IIA string can be derived from the supermembrane by a 
``double dimensional reduction''.\cite{Duff:1987bx} 
For this purpose, one compactifies the $X^9$-dimension on
a circle, letting the membrane wind around this dimension by 
making the identification
$$\sigma^2 = X^9.$$
When the circle is shrunk to a point, the $\sigma^2$ is gone as well and
one is left with a string theory in ten dimensions. The resulting
Hamiltonian is just the well known Green-Schwarz Hamiltonian 
$$
{\cal L} = \frac1{2P^+_0}\Big(\vec P (\sigma)^2 + \vec X' (\sigma)^2\Big) 
-  \bar \theta \Gamma^- \Gamma^9 \theta' (\sigma)
$$
This is as expected if the supermembrane theory is a part 
of  M-Theory, one of whose defining properties is that it must
reduce to type IIA string theory upon compactification on a small circle. 
We would like to stress, however, that these considerations are by
no means sufficient to establish that superstrings are contained
in the supermembrane at the dynamical level. To recover the full 
dynamics of superstring scattering amplitudes, and in particular the 
full multistring vertex operators, is quite a different, and far 
more complicated task (see e.g. H. Verlinde's lectures on matrix string 
theory at this School).

While these properties are for the most part almost self-evident, 
and also in accord with everything we know from superstring 
theory, we now return to the degeneracies of the Hamiltonian
pointed out above. These constitute a very important physical property 
of the membrane, whose full significance has come to be fully 
appreciated only more recently. Namely, the spikes described 
above causing the instabilities need not have ``free ends''. 
It can also happen that a membrane quenches and separates into 
several parts that are connected by infinitesimally
thin tubes, \psfig{Three membranes connected by tubes}{multimem\eps}
see Fig.\ \the\figcount.  Because these tubes do not carry any energy,
such a configuration is physically indistinguishable from the
multi-membrane configuration obtained by removing the strings
connecting the various pieces. In this way, membranes are allowed to
split and merge, and there is no conserved ``membrane number''.
In fact, the generic configuration occurring in a path integral
formulation will be such that we cannot even assign a definite 
``membrane number'' to a given configuration.
  
Similar remarks apply to the treatment of different membrane 
topologies. For instance, one can build a toroidal membrane
from a spherical membrane by connecting two points by a stringlike
tube without any additional cost in energy. Hence it does not really
make sense to talk about membranes of fixed 
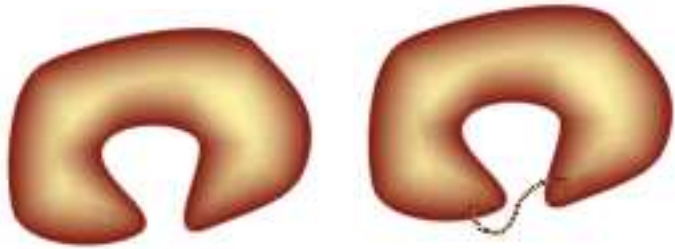
topology \psfigs{Equivalent membranes with different
topologies}{topchange1\eps}{topchange2\eps}, see Fig.\ \the\figcount.

This new interpretation represents an important change of viewpoint
vis-\`a-vis the one widely accepted a few years ago, when people were
trying to construct a first quantized version of the supermembrane. 
The futility of these attempts became glaringly obvious
when it was shown that the spectrum of the ($SU(N)$ version of the)
Hamiltonian \lcham\ is continuous.\cite{deWit:1989ct} This
looked like (and in fact was) a disaster for the interpretation
of the quantum supermembrane as a first quantized theory, and prompted 
several attempts to argue the continuous spectrum away.  
We note that, of course, the instabilities would go away if 
the effective action of the quantum supermembrane contained terms 
depending on the world volume curvature, which would suppress spikes 
and stringlike tubes. However, such terms presumably would not be 
compatible with the (desired) renormalizability of the theory, but 
would also spoil the correspondence with the matrix model.

To summarize: the old objection that membrane theories {\em ``can't be
first quantized''} --- meaning that the Hamiltonian is not quadratic
unlike the superstring Hamiltonian --- acquires a new and much deeper
significance: the theory cannot be first quantized {\em because the
quantum supermembrane is a second quantized theory from the very
beginning!} This shows why membrane theory is so hard: we are dealing
with a theory where the very notion of a one- (or multi-) particle
state can only be extracted in certain asymptotic regimes, if it makes
sense at all.

\chapter{Area Preserving Diffeomorphisms}
We now return to the canonical constraint \sconst\ which expresses
the invariance of the lightcone supermembrane under area preserving 
diffeomorphisms (or just APD, for short). To further analyze 
this residual symmetry, we introduce the following bracket on the
space of functions $A, B$ of the membrane coordinates 
$$
\{A,B\} (\sigma):= {\epsilon^{rs}\over\w}\d r A(\sigma)\d s B(\sigma)
$$
This is the Poisson bracket associated with the symplectic form 
$$
\omega = {\epsilon_{rs}\over\w} d\sigma^r \wedge d\sigma^s
$$
which, as any two form in two dimensions, is closed. The bracket
is manifestly antisymmetric
$$
\{A,B\} = - \{B,A\}
$$
and obeys the Jacobi identity
$$
\{A,\{B,C\}\}+\{C,\{A,B\}\}+\{B,\{C,A\}\}=0.
$$
and therefore has all the requisite properties of a Lie bracket.
Consequently the functions on the membrane naturally form an infinite
dimensional Lie algebra with the above bracket. The dimension
$p=2$ of the membrane is essential here, as these
statements have no analog for $p>2$ (at least not within the
framework of ordinary Lie algebra theory). 

By means of this bracket we can rewrite the potential density as
$$
\bar g = \det (\d r\vec X\cdot \d s\vec X)= (\{X^a,X^b\})^2
$$
Similarly, the non-zero mode part of the Hamiltonian \lcham\ can be
cast into the form
$$\eqalign{
{\cal M}^2 &= 
 \int d^2\sigma\biggl[{1\over 2\w}\biggl[\vec
P^2(\sigma)\biggr]^\prime
+ \w \biggl(\br 12\{\X a,\X b\}^2
+\bar\O\Gamma^-\Gamma^a
\{\X a,\O\}\biggr)\biggr]\cr}    \gln\mass
$$
Again, the prime indicates that the zero modes have been left out. 
The advantage of rewriting the previous formulas in this peculiar 
way is that the similarities with Yang Mills theories
become quite evident: the potential 
energy vanishes whenever $\{\X a,\X b\} =0$, which is equivalent
to the statement that the $\X a$ belong to the Cartan subalgebra
of the Lie algebra introduced above. When we truncate 
this infinite dimensional Lie algebra to a finite dimensional 
matrix Lie algebra, the zero energy configurations just correspond 
to the diagonal matrices.

Because
$$
\{\X a,\X b\}={\epsilon^{rs}\over\w}\d r\X a\d s\X b
$$
is nothing but the volume (or rather: area) element of the membrane 
pulled back into space time, the residual invariance consists
precisely of those diffeomorphisms which leave the area density
invariant. These correspond to diffeomorphisms
$$
\sigma^r\mapsto\sigma^r+\xi^r(\sigma)
$$
generated by divergence free vector fields
$$
\d r(\w \xi^r(\sigma))=0.
$$
This equation is solved locally by 
$$
\xi^r(\sigma)= {\epsilon^{rs}\over\w}\d s\xi(\sigma)
$$
For topologically non-trivial membranes, there will be further
vector fields which cannot be expressed in this fashion, and which
correspond to the harmonic one forms in the standard Hodge decomposition.

As a little exercise readers may check that the commutator of two
APD vector fields is in one-to-one correspondence with the above 
Lie bracket in the sense that
$$
[\xi_1^ r\d r,\xi_2^s\d s]\longleftrightarrow\{\xi_1,\xi_2\}
$$
This means that the above Lie algebra can be identified with the Lie
algebra of divergence free vector fields which is the Lie algebra of
area preserving diffeomorphisms. Using this correspondence, we can 
calculate the variation of any function $f$ under an area
preserving diffeomorphism as
$$\delta f=-\xi^r\d rf=\{\xi,f\}.$$
It is now straightforward to verify that the mass $\cal M$ commutes 
indeed with the constraint generator \sconst. The latter can be 
reexpressed by means of the above Lie bracket as 
$$
\phi(\sigma)\equiv    
 %\epsilon^{rs}(\d r\vec P\d s\vec X-\d r\bar\O\gamma_-\d s\O)
\{ \vec P , \vec X \} - \{ \bar \theta , \Gamma^- \theta \}
\approx 0
$$
This should be compared to closed string theory after the light cone gauge
has been imposed: In that case, the length preserving diffeomorphisms
are just the constant shifts
$$
\sigma\mapsto \sigma+ const.
$$
This constraint implies that the oscillator levels of the
left and the right movers are equal on physical states:
$$N_L=N_R$$
As is well known, already this single constraint has many non-trivial 
consequences!

\chapter{Supercharges and Superalgebra}
The supercharges expessing the global supersymmetry of the model are 
obtained by integrating the supercurrent \current\ over the membrane:
$$
Q = \int d^2\sigma J^0 \equiv Q^+ + Q^-
$$
The chiral supercharges are given by
$$
\eqalign{Q^+&\equiv\br 12\Gamma^+ \Gamma^- Q=\int
d^2\sigma\left(2P^a\Gamma_a+\w\{\X a,\X b\}\Gamma_{ab}\right)\O\cr
Q^-&\equiv\br 12\Gamma^- \Gamma^+ Q = \int d^2\sigma S=2\Gamma^-\O_0\cr }
$$
We see that $Q^-$ acts only on the fermionic zero modes. 
The zero mode part of $Q^+$ 
$$
Q^+_0 = 2P_0^a\Gamma_a\O_0 \equiv
Q^+ - Q^+_1
$$
is also conserved. Evidently, $Q_0^+$ is only relevant to the
center of mass kinematics.

With the help of the canonical Dirac bracket \canbrack\ one can now  
determine the full superalgebra. This calculation, including possible
surface contributions which could give rise to central charges was already
performed in \cite{deWit:1988ig}. The most general $d=11$ superalgebra 
has the form \cite{Townsend:1997wg}
$$
\{Q_\alpha ,\bar Q_\beta \} = 
     \Gamma^\mu_{\alpha\beta} P_\mu 
  +  \br 12 \Gamma^{\mu\nu}_{\alpha\beta} Z_{\mu\nu} +
 \br 1{5!}\Gamma^{\mu\nu\rho\sigma\tau}_{\alpha\beta} Z_{\mu\nu\rho\sigma\tau}.
$$
The corresponding Lie-superalgebra, including the bosonic commutation
relations of the bosonic operators, is known as $OSp(1|32)$ in
mathematical terminology. We recognize that besides the membrane
charge $Z_{\mu\nu}$ the algebra also admits a 5-brane charge
$Z_{\mu\nu\rho\sigma\tau}$. However, a recent investigation of the
lightcone superalgebra for the supermembranes with winding
\cite{deWit:1997fp,deWit:1997jz,deWit:1997zq,Ezawa:1997hq,Banks:1997my}
has shown that this 5-brane charge is, in fact, absent despite the
fact that $d=11$ supergravity admits both 2-brane and 5-brane-like
solutions.\cite{Duff:1991xz,Gueven:1992hh} This little puzzle
should not come as a total surprise in view of the following fact.
While the 2-brane couples to the 3-index field of 11d supergravity,
the 5-brane charge would couple to a ``dual'' 6-index field; one would
therefore expect the 5-brane charge to be related to another version
of 11d supergravity with a 6-index field. However, for all we know
such a version of 11d supergravity does not
exist,\cite{Nicolai:1981kb} but see \cite{Bandos:1997gd} for 
more recent references and a reformulation containing both a 3-index 
$and$ a 6-index tensor field).

The remaining part $Q_1^+$ of $Q^+$ obtained upon removing all 
zero-mode contributions contains the non-trivial information about 
the supermembrane dynamics. It gives rise to the superalgebra
$$
\{Q_{1\alpha}^+,\bar Q_{1\beta}^+\} =
  \Gamma^+_{\alpha\beta}{\cal M}^2
$$ 
The existence of massless states at threshold is equivalent to finding a 
normalizable state $\Psi_0$ obeying ${\cal M}^2 \Psi_0 =0$. Once such 
a state is found, a whole massless multiplet of $d=11$ supergravity 
is generated by acting with the zero mode supercharges given above.
If ${\cal M}^2$ does not annihilate the state, one gets many more
states, and the supermultiplet is a massive (long) supermultiplet
of $d=11$ supersymmetry. If $d=11$ supergravity is to emerge as a low
energy limit from the supermembrane there must be normalizable states 
that are annihilated by the quantum operator for ${\cal M}^2$.
Let us also note that the winding states of minimal mass are BPS states, and 
hence belong to short multiplets, too \cite{deWit:1997zq,Ezawa:1997hq}.

% Including matrix.tex
\chapter{Supermembranes and Matrix Models}
We are now ready to establish the connection between the supermembrane
Hamiltonian and the large $N$ limit of supersymmetric $SU(N)$ matrix
model,\cite{deWit:1988ig} which also underlies the recent proposal for
a concrete formulation of M-Theory.\cite{Banks:1997vh} To
this aim, we need to truncate the supermembrane theory to a
supersymmetric matrix model with finitely many degrees of freedom. The
idea is then to {\em define} the quantum supermembrane as the limit
where the truncation is removed, taking into account possible
renormalizations.  The truncated model can be alternatively obtained
by dimensional reduction of the maximally supersymmetric $SU(N)$ Yang
Mills theory from 1+9 to 1+0 dimensions, a reduction which had been
originally investigated in
\cite{Claudson:1985th,Baake:1985ie,Flume:1985mn}).  Upon quantization
it becomes a model of supersymmetric quantum mechanics with extended
(${\cal N}=16$) supersymmetry.

To proceed, we expand all superspace coordinates in terms of some
complete orthonormal set of functions $Y_A(\sigma)$ on the membrane
$$
\vec X(\sigma) = \vec X_0 +\sum_A \vec\X A Y_A(\sigma)
$$
We next define a metric on this function space by
$$
\int d^2 \sigma\w Y_A(\sigma)Y_B(\sigma)=\eta_{AB}.
$$
by means of which indices $A,B,\ldots$ can be raised and lowered,
such that we have the orthogonality relations
$$
\int d^2 \sigma\w Y_A(\sigma)Y^B(\sigma)=\delta_A^B.
$$
For instance, for spherical membranes, we can take for the $Y^A$
the standard spherical harmonics $Y_{lm}(\theta,\varphi)$, in which case 
raising and lowering corresponds to complex conjugation such that
$Y^{lm}= (Y_{lm})^* = Y_{l,-m}$. For a real basis of orthogonal
functions, we can choose $\eta_{AB}=\delta_{AB}$.
For the toroidal membrane, the indices are of the form $A=(m_1,m_2)$, 
where each index labels a Fourier mode. It is convenient but not 
strictly necessary to take the $Y_A$ to be eigenfunctions of the 
corresponding Laplace-Beltrami operator (w.r.t. the background
metric $w_{rs}$). We also record the general completeness relation
$$
\sum_AY^A(\sigma)Y_A(\sigma')=\br 1\w \delta(\sigma-\sigma')
$$

Because of the completeness relation we can express the Lie
bracket in terms of the new basis:
$$
\{Y_A,Y_B\}=\f ABC Y_C
$$
One finds the structure constants to be 
$$
\f ABC = \int d^2\sigma\, \epsilon^{rs}\d rY_A\d sY_B Y^C.
$$

Exploiting these observations we now wish to truncate the theory to
another one with only finitely many degrees of freedom. This 
``regularization'' of the supermembrane is achieved by introducing a 
cut-off on the number of modes, such that the mode indices 
$A,B \ldots$ are restricted to a finite range of values
$1,\ldots, \Lambda$. Consistency then demands that the group of 
APDs must be approximable by a finite Lie group $G_\Lambda$ with
${\rm dim}\, G_\Lambda = \Lambda$ in the sense that
$$
\lim_{\Lambda\rightarrow\infty} \f ABC \big( G_\Lambda \big) = \f ABC
({\rm APD})
$$
for any fixed triple $A,B,C$. The crucial result 
$$
G_\Lambda = SU(N) \qquad {\rm with} \qquad \Lambda = N^2 -1
\gln\SUN
$$
was first established for spherical membranes in \cite{Goldstone,Hoppe}. 
Later on, it was extended to toroidal
membranes \cite{torus1,torus2,deWit:1990vb}, 
and finally to membranes of arbitrary genus\cite{Bordemann:1994zv}. 
The APD Lie algebra is thereby replaced
by a finite dimensional Lie algebra, namely the Lie algebra of $SU(N)$
matrices.  To emphasize the matrix character of this regularization,
we will thus replace the Lie bracket by a commutator:
$$
\{\cdot,\cdot\}\longrightarrow[\cdot,\cdot].
$$ 
Before continuing, we would like to point out one essential
difference between this ``regularization'' and the lattice 
regularization of gauge theories, which look very similar at 
first sight: Unlike in lattice QCD, there is no ``small parameter'' 
here analogous to the lattice spacing (or the inverse momentum cutoff).   
To see this, we note that large $N$ does not necessarily mean 
``large energy'', because for any given $N$ we can find configurations 
of arbitrarily small energy by taking commuting $N \times N$ matrices.

Although there is no room here to describe the construction in 
full detail, we would like to give readers at least an idea
how the correspondence between $SU(N)$ and APD works for toroidal 
membranes. This is the simplest case because we can make use 
of a double Fourier expansion. Choosing coordinates
$0\leq \sigma_1, \sigma_2 < 2\pi$, we have the orthonormal basis
$$
Y_{\vec m}(\vec\sigma) = 
 {1\over{\sqrt{4\pi}}} e^{i\vec m\cdot\vec\sigma}
$$
The APD structure constants follow from
$$
\{Y_{\vec m},Y_{\vec n}\} = -4\pi^2 (\vec m\times\vec n)Y_{\vec
m+\vec n}.\gln\ylie
$$
To see the relation with $SU(N)$, we employ the 't-Hooft clock and
shift matrices 
$$\eqalign{
U&=\pmatrix{0&1&&\cr&\ddots&\ddots&\cr &&\ddots&1\cr 1&&&0\cr}\qquad
V=\pmatrix{1&&&\cr&\omega&&\cr&&\ddots&\cr&&&\omega^{N-1}\cr}\cr}
$$
where $\omega$ is an $N^{\rm th}$ root of unity $e^{2\pi
ik/N}$. They commute up to a phase factor:
$$UV=\omega VU$$
Any traceless $N\times N$ matrix can be written as a linear
combination of matrices $U^{m_1}V^{m_2}$. One finds the matrix
commutator to be
$$\eqalign{
&\left[U^{m_1}V^{m_2},U^{n_1}V^{n_2}\right]=
(\omega^{m_2n_1}-\omega^{m_1n_2})
U^{m_1+n_1}V^{m_2+n_2}\cr}
$$
If we now take $N$ to infinity keeping $\vec m$ and $\vec n$ fixed
this approaches
$$\eqalign{
&\lim_{N\to\infty}\left[U^{m_1}V^{m_2},U^{n_1}V^{n_2}\right]\to
{2\pi ik\over N} (\vec m\times\vec n)U^{m_1+n_1}V^{m_2+n_2}.\cr}
$$
So, in this limit this is the same Lie-algebra as
the one we found for the $Y_{\vec m}$ in \ylie. This corroborates
our claim that the Lie algebra of area preserving diffeomorphisms 
on the torus can be approximated by $su(N)$.

At this point we should emphasize that the question of how to
rigorously define the notion of ``limit'' here is very subtle.
Namely, the above statements hold only for particular 
bases of $SU(N)$ matrices, which must be specially and 
differently chosen for every given membrane topology. 
In other words, the limits are highly basis dependent.
This explains the --- at first sight paradoxical --- fact that the
area preserving diffeomorphisms for {\em different} membrane
topologies can be approximated by the {\em same} group: while all
bases are equivalent for finite $N$, this is no longer true in
the limit $N\rightarrow \infty$ because the large $N$ limit of 
the corresponding equivalence transformations will not exist.
So for instance we have ${\rm Diff}_0 (S^2)\neq {\rm Diff}_0 (T^2)$,
and neither of the associated Lie algebras is isomorphic to
$su(\infty)$ (defined as the set of $\infty \times \infty$ matrices
with only finitely many non-vanishing entries), see e.g.
\cite{BHMS,Bordemann:1994zv}. Observe also that the APDs generated
by harmonic vector fields (whose number depends on the membrane
topology) have no finite $N$ analogs. On the other hand, in the
quantum theory, where we are mainly interested in the large $N$ limit 
of gauge invariant correlators, and not in approximating $C^\infty$ 
diffeomorphisms, these issues are no longer so prominent.

Our main point here is that the restriction to finite $N$ should be
regarded as a non-perturbative regularization that can be used
to give a rigorous and non-perturbative {definition} of the 
quantum supermembrane.\footnote{However, it is conceivable that the 
quantum mechanical model with infinite dimensional gauge group APD
can be made sense of {\em eo ipso}, in which case we cannot even
rule out the possibility that it is actually
different from the large $N$ limit of the matrix model.} This 
interpetation is similar in spirit to lattice gauge theory whose 
continuum limit is {\em by definition} the quantum Yang Mills theory 
(provided the limit exists). We have seen in the above discussion that 
this quantum matrix model in principle describes membranes 
of all possible topologies in the large $N$ limit. 

After these preliminary remarks, we can now write down the truncated 
Hamiltonian. Before doing so, it is convenient to switch from
the $SO(1,10)$ basis to a basis of $SO(9)$ $\gamma$ matrices,
as this is the residual symmetry of the lightcone gauge theory.
The precise relation is
$$
\Gamma^+ = \pmatrix{0 & \sqrt{2} \cr 0 & 0 \cr}\otimes\ID \qquad
\Gamma^- = \pmatrix{0& 0 \cr \sqrt{2} & 0 \cr}\otimes\ID 
$$
and 
$$
\Gamma^a = \pmatrix{0 & 1 \cr 1 & 0 \cr} \otimes \gamma^a
$$ 
where the $\gamma^a$ are the standard $SO(9)$ $\gamma$-matrices.
Accordingly, we will henceforth work with 16 component (real)
spinors of $SO(9)$, eliminating all $\Gamma^\pm$ from the equations.
As we already mentioned, the matrix Hamiltonian is then
no\-thing but the dimensional reduction of the Hamiltonian
of maximally extended super Yang Mills theory with gauge group
$SU(N)$ from 9+1 down to 0+1 dimensions:
$$
\eqalign{ H&=\br 12 P_a^AP_{aA} + \br 14 (\f ABC X_a^AX_b^B)^2-
\br i2f_{ABC}X_a^A\O^B\gamma^a\O^C\cr}\gln\mham
$$
In writing this Hamiltonian, we have suppressed the Yang Mills 
coupling constant which can be identified with the longitudinal
momentum $P^+_0$, see \lcham. The Gauss constraint reads
$$
\phi_A=f_{ABC}(X_a^BP_a^C-\br i2 \O_\alpha^B\O_\alpha^C)\approx 0
$$
now simply expresses the $SU(N)$ gauge invariance of the model.
It is straightforward to check that it commutes with the Hamiltonian, i.e.
$$
\{ \phi^A , H \} = 0
$$
The global supersymmetry is reflected in the
existence of the supercharges
$$
Q_\alpha= ( P^a_A\gamma_a+\br
12f_{ABC}X_a^BX_b^C\gamma^{ab})_{\alpha\beta}\O_\beta^A.
$$
As usual, the superalgebra relation
$$
\{Q_\alpha,Q_\beta\}\approx 2\delta_{\alpha\beta}H
$$
implies the lower boundedness
$$H\ge 0.$$
of the energy spectrum.

\chapter{Spectrum}
The supersymmetric Hamiltonian \mham\ has a continuous spectrum,
unlike the bosonic Hamiltonian contained in it. We will illustrate the
idea of the proof by means
of a simpler toy model. The proof in this case is much simpler but, 
apart from technicalities, the same as for the supermembrane. The model 
is a two dimensional supersymmetric quantum mechanical system 
with flat valleys.  The supercharges are  
$$
Q=Q^+=\pmatrix{-xy&i\d x+\d y\cr i\d x-\d y&xy.\cr}
$$
They square to yield the Hamiltonian
$$
H=\br 12\{Q,Q^+\}=\pmatrix{-\Delta+x^2y^2&x+iy\cr x-iy&-\Delta+x^2y^2\cr}
$$
which acts on two component wave functions
$$
\Psi =\pmatrix{\psi_1(x,y)\cr\psi_2(x,y)\cr}.
$$
The potential $V(x,y)=x^2y^2$ has valleys along the coordinate axes.
\psfig{The potential of the toy model}{valey\eps} 

Consider, for example, the point $(x,0)$
for some large $x$. If we take into account only the bosonic part given by the
diagonal values of the Hamiltonian, we would find a harmonic oscillator
potential in the $y$ direction with frequency proportional to $x$. The
zero point fluctuations therefore induce an effective $|x|$ potential
for the slower motion in the $x$ direction which confines the wave
function and prevents it from leaking out to $x\to\pm\infty$. So the
Born-Oppenheimer approximation that is valid for large values of $x$
lets us expect a normalized ground state with finite $E>0$ which 
is located near the origin (nowadays also called ``the stadium'').  
Because of this confinement the spectrum is discrete despite the
presence of flat directions.\cite{Simon:1983jy,Luscher:1983ma}

To make these qualitative considerations a little more precise,
we write the bosonic Hamiltonian as
$$\eqalign{H_B&= \br 12(p_x^2+p_y^2)+\br 12 (p_x^2+x^2y^2)
+ \br
12(p_y^2+x^2y^2)\cr
&\ge \br 12 (p_x^2+p_y^2)+\br 12 |y|+\br 12|x|\cr}$$
The inequality is meant as an inequality of operators. Take the
expectation value of the bosonic Hamiltonian in any state: It is
bounded from below by the expectation value of the second operator. 
But for the latter, the valleys have disappeared and we are left
with a confining potential (which looks like the inverted pyramid
in the Paris Louvre). Therefore, there can only be a finite number 
of states with an energy less than this expectation value. Thus we
have shown that the spectrum of the bosonic model is indeed discrete.

But things change dramatically when supersymmetry is turned on: Now,
the motion in the $y$ direction is described by the supersymmetric
analogue of a harmonic oscillator whose bosonic zero point
energy is cancelled by the fermionic contribution. Thus, there is no
confining potential any more and the wave function can escape to
infinity. To give a rigorous proof of this fact for any given energy
$E\geq 0$, we will construct a trial wave function $\Psi$ such that
$\|(H-E)\Psi\|$ is less than any given positive number.

To this aim, we first define a function $\chi(x)$ of one real
variable for any given $E\geq 0$ as follows:
$$
\chi (x) := e^{ikx} \chi_0 (x)
$$
where $k=\sqrt{E}$ and $\chi_0$ is a slowly varying real function of
compact support in $\MR$ normalized such that
$$
\|\chi\|^2 = \int\!\chi^2dx = 1 
$$
Observe that $\chi_0$ can be chosen in such a way that its derivatives 
can become arbitrarily small while it is still normalized to one.
Then we define, for $\lambda\in\MR$,
$$
\Psi_\lambda(x,y) := \chi(x-\lambda)\sqrt{|x|\over 4\pi}e^{-\br
12|x|y^2} \pmatrix{1\cr-1\cr}.  \gln\Psil
$$
This is a harmonic oscillator wave function in the variable $y$
transversal to the valley along the $x$ axis, but with a frequency
proportional to $x$, and judiciously chosen such that the bosonic and
fermionic zero point energies cancel to better and better accuracy
as we push this wave fucntion further into the valley by picking 
$\lambda$ sufficiently large. In this way, given any $\epsilon>0$, 
we can find $\lambda$ such that 
$$
\|(H-E)\Psi_\lambda\|<\epsilon,
$$
This proves that the spectrum of $H$ consists of all non-negative numbers. 

The results on the continuity of the spectrum for the supermembrane
are entirely analogous.\cite{deWit:1989ct} Again, the $SU(N)$
truncated bosonic membrane has a discrete spectrum, but the spectrum
of its supersymmetric extension is continuous, extending all the way
down to $E=0$. This result is quite contrary to what one would expect
from the superstring, whose discrete excitations are interpreted as
one-particle excitations of a higher dimensional target space theory.
Had we expected the model to describe a first quantized membrane, this
would have been the end of the story. With this interpretation, it
would be impossible to find particle-like excitations in the low
energy theory the way it is possible in string theory.

However, the result acquires a completely different significance
in view of the multi-particle interpretation of the dimensionally
reduced super-Yang-Mills Hamiltonian \cite{Witten:1996im,Banks:1997vh} 
and our previous remarks in section 5. Namely, the continuous spectrum 
is now {\em required} for consistency, since scattering states of 
(super)membranes connected by tubes should come with a continuum 
of energies. Turning the argument around, we can now even claim that 
the discreteness of the bosonic membrane spectrum indicates the 
inconsistency of the quantized bosonic membrane! In order to perform 
the scattering calculations just mentioned to leading order, it is 
fortunately not necessary to know the detailed structure of the 
ground state: It is sufficient to know the asymptotic state when the 
two gravitons are largely separated, in which case the Born-Oppenheimer 
approximation becomes better and better. \cite{Plefka:1997hm}

\chapter{Renormalizability vs. $N\rightarrow\infty$ Limit}
The key question is whether the $N\rightarrow\infty$ limit of the 
supersymmetric matrix model exists. As we emphasized already, this is 
related to the question of whether the supermembrane makes sense as a 
full fledged quantum field theory, and whether it is renormalizable 
or even finite as a 2+1 dimensional quantum field theory.
A priori there is, of course, no reason to expect a complicated
Lagrangian such as \lagrangian\ to be so well-behaved. Indeed, the 
bosonic membrane is not renormalizable and thus most likely not 
viable as a quantum theory. However, maximal supersymmetry
could make all the difference here!

Treating the $X$'s and the $\O$'s as quantum fields on the world
volume, one can in principle calculate Feynman loop diagrams.  For the
supermembrane this has not yet been done, not least because of
technical problems such as the absence of an off-shell formulation,
which makes any calculation extremely cumbersome. As in any quantum
field theory, the higher order amplitudes are potentially divergent.
This means that some regulator must be introduced, which possibly can
be removed only at the expense of certain renormalizations or higher
order counterterms respecting all symmetries. The only parameter
appearing in the supermembrane Lagrangian is the membrane
tension. Hence, introducing some cutoff $\Lambda$, we would set $T_2 =
T_2(\Lambda)$ and try to adjust this dependence in such a way that the
limit exists for all physically relevant correlators. The
supermembrane would be finite if all these limits existed with $T_2$
kept fixed.

If on the other hand the theory is non-renormalizable, the higher loop 
diagrams would necessitate an infinite number of counterterms of the form 
$R, R^2,\dots$, where $R$ is the world volume curvature tensor (actually
expressible through the Ricci tensor in three dimensions), just like
perturbatively treated matter coupled Einstein gravity in 2+1 dimensions. 
If this were the case, not only would the theory be ill-defined as a 
continuum theory, but also the connection with matrix theory would be lost: 
there seems to be no way to express the induced world volume 
curvature $R$ in terms of the APD Lie-bracket $\{X^a,X^b\}$ 
(for the membrane this was possible because the Lagrangian 
depends only on the metric determinant, not the curvature tensor). 

The hope is therefore that maximal supersymmetry is strong enough to
rule out possible counterterms for the $d=11$ supermembrane. Indeed,
to date no counterterm respecting the full ${\cal N}= 16$
supersymmetry and $\kappa$ symmetry is known (see e.g.
\cite{Paccanoni:1989hd} for an early discussion).
However, what is required to definitely settle this issue is an
off-shell formulation, or some equally powerful new mathematical tool
that would allow us to prove or disprove these statements. For the
time being, however, we have to be optimistic and simply assume that
the full $d=11$ supersymmetry does not admit higher order
counterterms. A further issue here is whether the regularization
procedure can be made to respect all symmetries.

How are these possibilities mirrored in the matrix model? Since the
finite $N$ matrix model is a perfectly well defined model of ordinary
quantum mechanics (and as such in principle easier to deal with than a
model of quantum field theory), the crucial question concerns the
$N\rightarrow\infty$ behavior. This limit is presumably different from
't Hooft's large $N$ limit with $g^2N$ kept fixed as
$N\rightarrow\infty$.  Moreover, there it is the $N=\infty$ theory
that is better defined (being based on planar diagrams only), whereas
it is the finite $N$ theory which is ``harder'', and one tries to work
one's way from $N=\infty$ to finite $N$ (with $N=3$ for QCD). Here, we
are trying to do the opposite.

The first possibility is that this limit simply does not exist, 
no matter how the parameters are varied. This might be due 
to large $N$ divergences, which would be analogous to the
divergences underlying the potential non-renormalizability of the
supermembrane. But it might also arise from a non-analytic
$N$ dependence even in the absence of genuine large $N$ divergences. 
We will present an example of such behavior in section~12. Again, since 
we expect the large $N$ matrix model to be a description of M-Theory, 
the non-existence of a suitable large $N$ limit would be the end for this 
conjecture. 

The second possibility is that a limit does exist if we vary and
renormalize the Yang-Mills coupling constant $g=g(N)$ as we go to 
large $N$. More precisely, the function $g=g(N)$ would have to be
{\em universal} in the sense that the limit
$$
\lim_{N\to\infty}{1\over f(N)}\Big\langle\ldots\Big\rangle_{g(N)}
$$
exists simultaneously for all physically relevant correlators.
Here, $f(N)$ is an appropriate wave function renormalization that 
might depend on the correlator under consideration. This means that 
the large $N$ limit would be a {\em weak limit}, similar to the 
asymptotic limit relating the in- and out-fields of LSZ quantum field 
theory to the interacting fields. This would be the matrix analog of 
renormalizability. Finally, if we could leave $g$ fixed independently 
of $N$, this would correspond to finiteness of the supermembrane. 
In any case, the question of how to properly define the limits 
relevant to the approximation of $C^\infty$ area preserving diffeomorphisms 
on membranes of different topology appears to be of no great 
relevance in the quantum mechanical context.

\chapter{Relation to D-Branes}
Over the last two years, the supersymmetric matrix model has attracted
a great deal of attention and stirred up quite some excitement.  The
proposal according to which the matrix model for $N\rightarrow\infty$
{\em is} M-Theory on a flat background\cite{Banks:1997vh}, had its
roots in the discovery of D-branes and their role in the description
of non-perturbative string excitations \cite{Polchinski:1995mt} and
the realization that the dynamics of these objects is governed by
dimensionally reduced Yang Mills theories.\cite{Witten:1996im} These
developments were independent of supermembrane theory, and it is
therefore all the more remarkable that one ends up with the same model:
$SU(\infty)$ super-Yang Mills theory reduced to one dimension. Here,
we would like to briefly review how the matrix model arose in the
context of D0 branes, referring readers to C.~Johnson's lectures at
this School for a more detailed treatment and many further references.

One distinction between the M(atrix) theory and the quantum 
supermembrane is the treatment of the longitudinal momentum: 
in the proposal of \cite{Banks:1997vh} it is identified with the 
number of longitudinal quanta for gauge group $SU(N)$:
$$
P^+_0=\frac{N}{R_s}
$$
where $R_s$ is the compactification radius, whereas in the supermembrane
the longitudinal momentum is treated as a canonical 
variable having non-trivial commutation relations. The above identification 
may be viewed as an indication that the matrix model is physically 
meaningful already for finite $N$, and should be identified 
with M-Theory compactified on a light-like circle.\cite{Susskind:1997cw} 
This proposal was made more precise in \cite{Seiberg:1997ad,Sen:1997we}: 
One should think of the light-like circle as a small space-like circle 
with radius $R_s$ that is boosted by a large amount. Then one should take
the $R_s$ to zero while compensating with further boosts in order to
keep the Hamiltonian finite. One finds that, in this limit, the ten
dimensional Planck mass goes to infinity.  But M-Theory, when
compactified on a small space-like circle, is (almost by definition)
type IIA string theory, when one identifies the Kaluza Klein
excitations with the non-perturbative (BPS type) string states. As
lightest solitonic objects, type IIA string theory contains
D0-branes. These are particles on which open strings can end. Since
the energies of the excited string modes scale with the Planck mass,
which goes to infinity, these excited modes become infinitely
heavy. Only the zero mode dynamics of the strings survive. 
Since the open strings are attached to the D0-branes the gauge fields
live only on the world lines of the D0-branes. If there are $N$ 
different D0-branes, strings can have $N$ different positions for 
each of their two end points. This is the D0-brane explanation for
the emergence of Chan-Paton labels, an the resulting theory is
ten dimensional $SU(N)$ super-Yang Mills theory dimensionally reduced
to one time dimension (the world line of the D0-branes). 

Unfortunately, the story is a bit more complicated than it seemed in
the beginning: It has been known for quite some time that the 
treatment of the zero modes is quite subtle when one attempts to 
compactify a light-like direction. Furthermore, the low energy limit 
of M-Theory that is used for this proposal, namely $d=11$ supergravity, 
is strictly valid only as long as the D-branes are far apart 
and the curvature is small. On the other hand, the regime in which 
the Yang-Mills description of the D-brane dynamics is valid is the 
limit where the D-branes approach each other at substringy distances.
The possibility that the supergravity description is also
valid for some short range processes was indicated by the agreement of
scattering amplitudes calculated in supergravity and in the
D-brane/matrix picture \cite{Douglas:1997yp,Becker:1997wh} for two
graviton scattering and more recently for three graviton scattering.
\cite{Okawa:1998pz} It has been shown in \cite{Paban:1998ea} 
that at least the two graviton term is protected by supersymmetry 
at all distance scales. 

There are also arguments, due to \cite{Douglas:1997uy}, that in curved 
supergravity backgrounds the truncation to finitely many degrees 
is not possible because the excited string modes do not decouple. 
But this problem might be overcome in the full supermembrane theory 
since it is also consistent for arbitrary backgrounds that satisfy 
the supergravity equations of motion. It its not at all clear how 
these backgrounds could be incorporated into the D0-brane picture,
although an obvious guess would be to repeat the procedure of
\cite{deWit:1988ig} to such a curved background.

M(atrix) theory naturally leads to a kind of non-commutative geometry
at short distances because the coordinates are represented by
non-commuting matrices. We noted earlier that the potential energy
vanishes when all the matrices commute with each other. But this means
we can diagonalize them simultaneously. So for zero energy the only
remaining degrees of freedom are the $N$ diagonal entries of the nine
matrices plus their fermionic partners. According to
\cite{Witten:1996im} we can interpret these eigenvalues as the
(commuting) coordinates of $N$ free particles moving in nine
dimensional transversal space.  The degrees of freedom corresponding
to the off-diagonal entries become heavy and can be integrated out
giving rise to effective interactions which resemble strings
stretching between the D0-particles.  The masses of these stretched
strings are proportional to the distance between the D0-branes. They
become important only when the particles are close together: only in
this case the off-diagonal matrix elements become light and the
matrices become non-commutative. On the other hand, this
non-commutativity means that the notion of positions makes less and
less sense for short distances. So M(atrix) theory can in principle
explain the emergence of a commutative space-time at long distances
(small energies) from a non-commutative ``spacetime foam'' at short
distances.

%, wheras at short distances it describes
%some kind of ``space-time foam''. This is in accord with what we would expect
%to happen in a full theory of quantum gravity and quantum space-time.

Finally, let us underline once more the multi-particle nature of the
quantum supermembrane. For finite $N$, a multi-membrane configuration
is approximated by a set of block diagonal matrices, while in the
D0-brane interpretation this matrix configuration describes
a bunch of coincident D0-particles. When the limit $N\rightarrow\infty$
is taken in such a way that each block becomes infinite dimensional,
each block matrix is thereby related to a membrane of its own.
The effective interactions between these blocks arise can then be computed 
by integrating over the off-diagonal elements. And indeed, the effective 
action at the one loop level can be interpreted as being due
to strings stretching between the different bunches of D0-branes, 
or alternatively between the separate membranes as in Fig.~3.
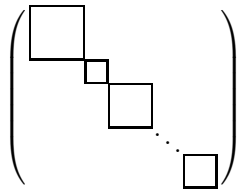
\def\luft{\kern 0.3ex}
$$\left(\luft\vcenter{
\offinterlineskip
\halign{#&#&#&\hss#\hss&#\cr
\hrulefill&\hss&\hss&&\hss\cr
\vrule\hbox to 20pt{\hss\vbox to 20pt{\vss}}\vrule&\hss&\hss&&\hss\cr
\hrulefill&\hss&\hss&&\hss\cr
\hss&\hrulefill&\hss&&\hss\cr
\hss&\vrule\hbox to 8pt{\hss\vbox to 8pt{\vss}}\vrule&\hss&&\hss\cr
\hss&\hrulefill&\hss&&\hss\cr
\hss&\hss&\hrulefill&&\hss\cr
\hss&\hss&\vrule\hbox to 16pt{\hss\vbox to 16pt{\vss}}\vrule&&\hss\cr
\hss&\hss&\hrulefill&&\hss\cr
\hss&\hss&\hss&$\vbox to 1em{\vss}\smash\ddots$\hss\cr
\hss&\hss&\hss&&\hrulefill\cr
\hss&\hss&\hss&&\vrule\hbox to 12pt{\hss\vbox to 12pt{\vss}}\vrule\cr
\hss&\hss&\hss&&\hrulefill\cr
}
}\luft\right)
$$

As we pointed out already, this multi-particle interpretation is now
also in complete accord with the continuous spectrum of the
matrix model Hamiltonian.

% Including lorentz.tex
\chapter{Light-cone gauge and Lorentz invariance}
The remaining two sections of these lectures are devoted to some
more specific technical issues, on which there has been
some progress recently. The first of these is the question 
of Lorentz invariance of the matrix model in the large $N$ limit.
As we will see the truncation to finite $N$ breaks the Lorentz 
invariance even at the classical level, but the symmetry is restored
as we take the limit $N\rightarrow\infty$.\cite{deWit:1990vb,Melosch,EMM}
The crucial question, which still awaits answer, is whether this remains 
true at the quantum level. Since the matrix model is well 
defined for any finite $N$, we can meaningfully address this 
problem at least in principle. 

Recall that after imposing light cone gauge we found an equation
\xminus\ for $X^-$ (where we have redefined the spinors by
factors of $w(\sigma)^{\frac14}$)
$$
\d rX^- = V_r = -{1\over P_0^+}\left(\br 1\w \vec P\cdot\d r\vec
X+\br i2\O\d r\O\right)
$$
We have to solve this differential equation and substitute the
solution into the expression for the Lorentz generators to check if
the algebra closes.

Following \cite{Goldstone,deWit:1990vb} we introduce a Greens function
and write
$$
X^-(\sigma) =\int d\sigma' \wpr
G^r(\sigma,\sigma')V_r(\sigma'),\gln\greens 
$$
always keeping in mind that this formula only works if $V_r$
obeys the integrability constraint $\partial_{[r} V_{s]}=0$. 
The requisite Greens function can be given rather explicitly.
Up to this point, the $Y^A$'s were an arbitrary complete set of 
orthonormal functions. We can now impose that they should
also be eigenfunctions of the Laplace-Beltrami operator
$$
\Delta =\br 1\w \d r (\w w^{rs}\d s)
$$
(recall that $w^{rs}$ is the inverse of the fiducial background
metric on the membrane)
$$
\Delta Y^A = -\omega^A Y^A
$$
On the complement of the constant mode $Y^0= const$, the operator 
$- \Delta$ is po\-si\-tive definite. The $\omega^A$ are therefore 
positive numbers. From the completeness relation
$$
\sum_A Y^A(\sigma) Y_A(\sigma')=\br 1\w\delta(\sigma,\sigma')
$$
we immediately see that the scalar Green's function 
$$
G(\sigma,\sigma') = -\sum_A {1\over\omega^A} Y^A(\sigma)Y_A(\sigma').
$$
satisfies
$$
\Delta G (\sigma,\sigma')=
{1\over\w}\delta^{(2)}(\sigma,\sigma') -1,\gln\deltagleichung
$$
the -1 being there because we have taken out the zero mode. 
The requisite Green's function is given by
$$
G^r(\sigma, \sigma') = \sum_A {1\over {\omega^A}} Y^A(\sigma)
w^{rs}(\sigma') \partial_s Y_A(\sigma')
$$

The above Green's function enables us to write down the explicit
expressions for the Lorentz generators in the lightcone gauge.  To do
so, one simply works out the Noether charges associated with the
original $SO(1,10)$ symmetry, and as in lightcone gauge string theory,
subsequently replaces the $X^-$ by \greens\ everywhere. This procedure
yields the following expressions\cite{deWit:1990vb}
$$
\eqalign{M^{ab}&= \int d^2\sigma(-P^aX^b+P^bX^a-\br i4\O \gamma
^{ab}\O)\cr
M^{+-} &= \int d^2\sigma (-P^+X^-+P^-X^+)\cr
M^{+a} &= \int d^2\sigma (-P^+X^a + P^a X^+)\cr
M^{-a} &= \int d^2\sigma (-P^-X^a+P^aX^-
{i\over
4P_0^+}\O\gamma^{ab}\O P_b-
{i\w\over 8
P_0^+}\{X_b,X_c\}\O\gamma^{abc}\O)\cr}
$$
When we substitute our expression for $X^-$ we find that $M^{+-}$ only
involves the zero modes. Expanding the expression for $M^{-a}$ one
encounters new overlap integrals that were not needed up to this point.
We have already made use of the APD structure constants
$$
f_{ABC}=\int d^2\sigma\sqrt{w} Y_A\{Y_B,Y_C\}
$$
In addition, we now need the symmetric invariants
$$
d_{ABC}=\int d^2\sigma\, \sqrt{w} Y_AY_BY_C
$$
which are APD analogs of the symmetric $SU(N)$ structure
constants 
$$\Tr(T^{(A}T^BT^{C)}).$$
 Both $f_{ABC}$ and $d_{ABC}$
are manifestly invariant under area preserving diffeomorphisms, 
because the overlap integrals only involve the metric determinant.
This corresponds to the invariance under $SU(N)$ of their
matrix analogs. However, for the full transcription of the 
lightcone gauge Lorentz generators into matrix language, we
need yet another overlap integral, viz.
$$
c_{ABC} = \int d^2\sigma \br\w{\omega^A} w^{rs}\d r Y_AY_B\d sY_C.
$$
To find its proper $SU(N)$ counterpart, we would have to introduce an
analog of the Laplace Beltrami operator on the $SU(N)$ Lie
algebra, such as for instance the expression 
$\sum_A [t^A,[t_A, \cdot]]$ (where the sum runs over a complete
basis of the Lie algebra); see e.g. \cite{HY,EMM}.
Another potentially worrisome feature is that $c_{ABC}$ depends
explicitly on the metric $w_{rs}$, and therefore is no longer
APD invariant. It is only by a ``miracle'' of supermembrane
theory that the concomitant ambiguities in the calculation of
Lorentz algebra will drop out at the end. This may for instance
be seen by working out the APD transformation on $X^-_A$, which 
also lacks manifest covariance \cite{deWit:1990vb}:
$$
\delta X^-_A = \xi^B \f ABC X^-_C + {\xi^B\over{2P^+_0}}
  (c_{ABC}+ c_{ACB})\phi^C
$$ 
where $\xi^A$ is are the orthonormal basis coefficients of
the APD transformation parameter. Happily, the non-covariant 
terms in this variation vanish precisely on the
physical subspace where $\phi^A\approx 0$.

As is well known from superstring theory (see e.g. \cite{Green:1987sp}), 
the crucial part in the check of Lorentz invariance is the
bracket of two boost generators: 
$$
\{M^{-a},M^{-b}\} \approx 0\gln\critcom
$$ 
which must vanish on the physical states. For the supermembrane,
this can be shown indeed after a lot of algebra.\cite{Melosch,EMM}
Let us just mention that in course of the calculation one has to employ 
identities such as
$$
\eqalign{f^E{}_{[AB}f_{C]DE}&=0\cr
f^E{}_{A(B}d_{CD)E}&=0\cr
d_{ABC}f^A{}_{[DE}f^B{}_{F]G}&=0\cr
c_{DE}{}^{[A}f^{BC]E}&=0\cr}
$$
These identities are valid for the APD tensors, but some of them are
violated in the finite $N$ truncation. The first is just the usual
Jacobi identity, which is clearly also valid for $SU(N)$. Likewise,
the second can be shown to hold also for the finite $N$ matrix
approximation. By contrast, the third holds only up to terms of order
$O(1/N^2)$; the fourth depends on what expression is used as the
finite $N$ analogue of $c_{ABC}$ and will be violated for any such
choice. This shows that the finite $N$ matrix model is not Lorentz
invariant, as was to be expected (this conclusion is also in line with
the interpretation of the finite $N$ model as a lightlike
compactification of M-Theory).  But let us emphasize again that at the
classical level Lorentz invariance is restored in the large $N$ limit,
where the identities are valid again.

String theory teaches us that the closure of the classical Lorentz
algebra by no means guarantees the Lorentz invariance of the quantum
theory. Quite to the contrary, (super)string theories are only quantum
consistent in certain critical dimensions where the anomalies
cancel. Let us recall the procedure there: one first defines the
physical Hilbert space by means of the Virasoro constraints. Then the
ordering ambiguities in the Lorentz generators are removed by
requiring that they  be well-defined, i.e. have finite matrix 
elements between any two physical states. This is achieved by
the normal ordering prescription. Only after this step does it make
sense to actually compute the algebra, and to determine the anomaly
--- with the well known result that the superstring can only live in
ten dimensions.

For the membrane Lorentz algebra, the situation is considerably more
complicated. The algebra of the $M^{\mu\nu}(N)$ for finite $N$ 
does not close even classically, as we just explained, but only in 
the limit $N\rightarrow\infty$. Replacing the classical phase space 
quantities by quantum operators, one immediately runs into quantum 
mechanical ordering ambiguities already for $N<\infty$ (where everything 
is still well defined). To proceed one would have to first establish the 
existence of the $N\to\infty$ limit, possibly with some $N$ dependent 
coupling $g(N)$. Besides the obvious ordering ambiguities, which may 
contribute ${\cal O}(N)$ or ${\cal O}(N^2)$ terms to the algebra and 
thereby yield unwanted extra contributions in the large $N$ limit, 
the construction of the quantum Lorentz generators may require 
extra renormalizations as is the case for composite operators 
in any renormalizable quantum field theory such as QCD. It would 
be very encouraging if one could show that this type of renormalization
was, in fact, unnecessary for the $d=11$ supermembrane. Remarkably, a
recent calculation \cite{Lowe} shows that the lowest non-trivial 
terms induced by quantum mechanical operator ordering do 
cancel! \footnote{Amazingly, this calculation seems to work for 
all classically allowed dimensions, so there is no indication of 
a critical dimension $d=11$ at the order considered.}

% Including massless.tex
\chapter{Massless states?}
Our second special topic where there has been considerable work
and some progress over the past two years concerns the question of 
massless states. Although there is still a scarcity of rigorous 
results for the $d=11$ supermembrane, there are hints that a 
normalizable groundstate wavefunction does exist, while none is 
expected to exist for the $d=4,5$ or $d=7$ supermembranes.

For pedagogical reasons (and because the relevant
equations are hardly ever written out in the literature!), we
we would first like to give some more details on the realization of 
the various operators. Straightforward replacement of the canonical 
brackets by quantum (anti)commutators immediately leads to
$$
\eqalign{[P_a^A,X_b^B]&=-i\delta_{ab}\delta^{AB}\cr
\{\O_\alpha^A,\O_\beta^B\}&=\delta_{\alpha\beta}\delta^{AB}\cr}
$$
A technical inconvenience is that the reality of the fermions
forces us to break the manifest $SO(9)$ invariance down to
$Spin(7)\times U(1)$, if we want an operator
realization of the fermionic brackets. This can be accomplished by
singling out  the 8 and 9 directions, and by combining the 16 real
$SO(9)$ spinors into two complex 8-component spinors, which are
complex conjugate to one another. Consequently, we set
$$
\eqalign{\lambda_\alpha^A :&=\br1{\sqrt{2}} (\O_\alpha^A+i\O_{\alpha+8}^A)\cr
\bar\lambda_\alpha^A :&=\br1{\sqrt{2}}
(\O_\alpha^A-i\O_{\alpha+8}^A)\equiv{\partial\over\partial\lambda_\alpha^A}
\cr}
$$
where $\alpha$ now runs over the values $1,\ldots, 8$ appropriate for
$Spin(7)$. The coordinates split into 
$$X_a^A =(X_i^A,Z^A,\bar Z^A)$$
with
$$\eqalign{Z^A := \br 1{\sqrt{2}} (X_8^A + iX_9^A)\cr
\bar Z^A := \br 1{\sqrt{2}} (X_8^A - iX_9^A)\cr}$$

The center of mass (zero mode) coordinates are split similarly.
As already explained, the zero mode part of the supercharge is completely 
separated from the (dynamical) rest of the theory. In the 
$Spin(7)\times U(1)$ the corresponding supercharges become
$$
\eqalign{Q_{0\alpha} &= -i\Gamma^i_{\alpha\beta}
{\partial\over\partial X_{0i}} \lambda_{0\beta}+ \sqrt 2
{\partial\over\partial Z_0} {\partial\over\partial\lambda_{0\alpha}}\cr
Q^+_{0\alpha}&= i\Gamma^i_{\alpha\beta}
{\partial\over\partial X_{0i}} {\partial\over\partial\lambda_{0\beta}}
+ \sqrt 2 {\partial\over\partial Z_0} \lambda_{0\alpha}\cr}
$$
They act on wave functions depending on the zero mode variables
$(X_0,Z_0,\bar Z_0)$ and $\lambda_{0\alpha}$. Since the fermionic
variables are treated as generating elements of a Grassmann algebra,
the wave function is a polynomial over this Grassmann algebra. 
The bosonic part of the wave function is made up of the even elements
$$
1\, ,\, \lambda_{0\alpha} \lambda_{0\beta}\, , \,
\lambda_{0\alpha} \lambda_{0\beta}\lambda_{0\gamma} \lambda_{0\delta},
\dots
$$
multiplied by plane wave functions of the bosonic coordinates
$(X_0,Z_0,\bar Z_0)$. Under $SO(9)$, these states transform as 
the ${\bf 44}\oplus {\bf 84}$ representations, corresponding to 
the on-shell degrees of freedom of the metric $g_{\mu\nu}$ and
the three index field $A_{\mu\nu\rho}$, respectively, of $d=11$ 
supergravity. The odd elements (again multiplied by plane wave functions)
$$
\lambda_{0\alpha}\, , \,
\lambda_{0\alpha} \lambda_{0\beta}\lambda_{0\gamma} \, , \,
\lambda_{0\alpha} \lambda_{0\beta}\lambda_{0\gamma} \lambda_{0\delta}
\lambda_{0\epsilon},
\dots
$$
transform as the $\bf 128$ (i.e. traceless vector spinor) of $SO(9)$ 
and thus comprise the on-shell degrees of freedom of the $d=11$ 
gravitino. Thus we recover precisely the representations corresponding 
to a massless multiplet of $d=11$ supergravity, i.e. the on-shell 
states of $d=11$ supergravity (also referred to as ``supergraviton'').

Readers should keep in mind that these considerations are purely
kinematical and by themselves provide no evidence for or against the
existence of a massless $d=11$ supermultiplet for either supermembrane
or M(atrix) Theory. In order to find the supergravity states in the 
Hilbert space of the supermembrane, one must identify the non-zero mode 
part of the wave function which is annihilated by the non-zero mode 
part of the full supercharge. Before we state what the problem is, 
it is instructive to have a look at the equations that must be 
solved for this purpose.

In the $Spin(7)\times U(1)$ formalism, the non-zero mode 
supercharges are given by \cite{deWit:1988ig}
$$
\eqalign{ Q_\alpha &=\biggl\{-i\Gamma_{\alpha\beta}^i
{\partial\over\partial X_i^A} + \br
12f_{ABC}X_i^BX_j^C\Gamma^{ij}_{\alpha\beta} - 
f_{ABC}Z^B\bar
Z^C\delta_{\alpha\beta} \biggr\}\lambda_\beta^A +\cr
&\quad+\sqrt 2\biggl\{ \delta_{\alpha\beta} {\partial\over\partial
Z^A} +
if_{ABC}X_i^B\bar Z^C\Gamma^i_{\alpha\beta}\biggr\}
{\partial\over\partial\lambda_{\beta A}},\cr
Q_\alpha^+ &=\biggl\{i\Gamma_{\alpha\beta}^i
{\partial\over\partial X_i^A} + \br
12f_{ABC}X_i^BX_j^C\Gamma^{ij}_{\alpha\beta}
+f_{ABC}Z^B\bar
Z^C\delta_{\alpha\beta} \biggr\} {\partial\over\partial\lambda_{\beta A}} +\cr
&\quad+\sqrt 2\biggl\{ -\delta_{\alpha\beta} {\partial\over\partial
\bar Z^A} 
+if_{ABC}X_i^B Z^C\Gamma^i_{\alpha\beta}\biggr\}
\lambda_\beta^A.\cr}
$$
The Gauss constraint is realized by the operator
$$\eqalign{
\phi^A &= f^{ABC}\biggl\{X_{iB} {\partial\over\partial X_{C}^C} + Z_B
{\partial\over\partial Z^C} +
\bar Z_B {\partial\over\partial\bar
Z^C}+\lambda_{\alpha B}{\partial\over\partial\lambda_\alpha^C}\biggr\}\cr}
$$ 
The supercharges satisfy the following anticommutation relations:
$$
\eqalign{\{Q_\alpha,Q_\beta\}&= 2\sqrt 2\delta_{\alpha\beta}\bar
Z^A\phi_A\approx 0\cr
\{Q_\alpha^+,Q_\beta^+\}&= 2\sqrt
2\delta_{\alpha\beta}Z^A\phi_a\approx 0\cr
\{Q_\alpha,Q_\beta^+\}&=
2\delta_{\alpha\beta}H-2i\Gamma^i_{\alpha\beta}X_i^A\phi_A \cr}
$$
The last anticommutator is thus weakly equal to the Hamiltonian
alias the mass operator ${\cal M}^2$.

The Hamiltonian can be split into two contributions, namely
$$
H=H_B+H_F.
$$
with the positive bosonic part
$$
\eqalign{H_B&=-\br 12{\partial^2\over\partial
X_i^A\partial X_i^A}-{\partial^2\over \partial Z^A\partial\bar Z^A}+\cr
&\qquad +\br 14\f ABE f_{CDE}\Bigl\{X_i^AX_j^BX_i^CX_j^D
+4X_i^AZ^BX_j^C\bar Z^D+2Z^A \bar Z^BZ^C\bar Z^D\Bigr\}\cr
&\ge 0\cr}
$$
and the fermionic part 
$$
\eqalign{H_F&=if_{ABC}X_i^A\lambda_\alpha^B\Gamma^i_{\alpha\beta}
{\partial\over\partial\lambda_{\beta C}}
+{1\over\sqrt
2}f_{ABC}\left\{Z^A\lambda_\alpha^B\lambda_\alpha^C -\bar Z^A
{\partial\over\partial\lambda_\alpha^B}
{\partial\over\partial\lambda_\alpha^c} \right\}.\cr}
$$
These operators act on the non-zero mode wave functions 
$$
\Psi = \sum_k\!\! \sum_{\alpha_1,\ldots,\alpha_k\atop A_1,\ldots,A_k}
\Phi^{\alpha_1\ldots\alpha_k}_{A1\ldots A_k}(x_i^A,Z^A,\bar Z^A) 
\lambda_{\alpha_1}^{A_1}\cdots\lambda_{\alpha_k}^{A_k}.
$$
where $k$ assumes only even values for bosonic wave functions; we
here anticipate that this wave function must be an $SO(9)$ singlet.
Since $\alpha_i \in \{1,\ldots,8\}$ and the $SU(N)$ indices take 
values in $\{1,\ldots,N^2-1\}$ these wave functions have 
$2^{8(N^2-1)}$ components. 

Let us now state the requirements that $\Psi$ must satisfy in order
to be an ac\-cep\-table non-zero mode wave function for  the groundstate
describing a massless state that combines with the zero mode wave function 
to yield a short (massless) $d=11$ supermultiplet.

First of all , $\Psi$ must be annihilated by the Hamiltonian 
or by the supercharges:
$$
H\Psi=0 \iff Q_\alpha\Psi=Q^+_\alpha\Psi=0  \gln\Grundzustand
$$
The equivalence of these equations is ensured by the physical
state constraints
$$\phi^A\Psi=0$$
for $A\in\{1,\ldots,N^2-1\}$. Inspection of the explicit operator
realization of the supercharges immediately shows that the
equations \Grundzustand\ relate all $2^{8(N^2-1)}$ components
of $\Psi$. Therefore we are dealing with a highly coupled system
of equations.

Secondly, the non-zero mode wave function must be a singlet of the 
rotational $SO(9)$. This is necessary because the zero mode part
of the full wave function by itself already provides the desired 
$SO(9)$ representations for $d=11$ supergravity, so that the 
remainder of the wave-function must obey
$$
J_{ab}\Psi\equiv (L_{ab} + S_{ab}) \Psi = 0
$$
where $L_{ab}$ and $S_{ab}$ are the ``orbital'' and ``spin'' parts
of the angular momentum operator, respectively. The singlet 
requirement is unique to eleven dimensions: in lower dimensions
the center of mass supercharges do not yield the full supergravity 
multiplet, and therefore the associated groundstate wave functions
could not be singlets.

Finally, $\Psi$ must be square integrable
$$
\|\Psi\|^2=\sum_k \sum_{\alpha_1,\ldots,\alpha_k\atop A_1,\ldots,A_k}
\|\Phi^{\alpha_1\ldots\alpha_k}_{A1\ldots A_k}\|^2<\infty
$$
where $\| \cdots \|$ denotes the usual $L^2$ norm. So altogether we 
must check the square integrability of $2^{8(N^2-1)}$ component 
functions! 

An important observation now is that, unlike in superstring 
theory, $\Psi$ {\em can not} be written as a tensor product
$$
\Psi = \Psi_B(X,Z,\bar Z) \otimes \Psi_F(\lambda,\bar\lambda),
$$
where both factors are separately invariant under $SO(9)$, i.e.
$L_{ab} \Psi_B = 0$ and $S_{ab} \Psi_F = 0$. To see this, recall 
that for the superstring, the supercharges are
of the form\cite{Green:1987sp} 
$$
Q_\alpha \propto \sum_{n\ne 0}\alpha_{-n}^i\Gamma^i_{\alpha\beta}
S_{n\beta}.
$$ 
The associated groundstate wavefunction is
$$
\Psi(\hbox{superstring})\propto \left(\prod_{n\ge 1} e^{-n\vec
X^2_n}\right) \left(\prod_{n,\alpha} \lambda_{n\alpha}\right)\gln\psistring
$$
with a somewhat schematic, but hopefully self-explanatory, notation:
the Dirac sea is ``half filled'' such that the state is either
annihilated by the bosonic oscillators (for $n<0$), or by the
fermionic oscillators (for $n>0$).  Both factors in \psistring\ are
obviously $SO(8)$ singlets.  To see that this factorization does not
work for a normalizable supermembrane (or matrix model) groundstate,
assume otherwise.  Then, since $H_F$ is linear in the coordinate
variables, its expectation value $(\Psi,H_F\Psi)$ contains a factor
$(\Psi_B, X^A_i \Psi_B)$ which vanishes by the assumed $SO(9)$
invariance of $\Psi_B$.  Consequently, we would also have
$$
(\Psi, H_F \Psi) = 0
$$
whence
$$
(\Psi, H\Psi)=(\Psi,H_B\Psi)=0
$$ 
But $H_B$ is a positive operator, and together with the assumed
square integrability of $\Psi_B$ this implies the vanishing of
$\Psi_B$, and hence of $\Psi$, which is a contradiction.
(If we did not assume normalizability of $\Psi_B$ here, we would
not be able to conclude $\Psi_B=0$ from this argument).

This means that $\Psi$ is annihilated by the total angular momentum
operators $J_{ab}$ that generate the $SO(9)$ rotations but the orbit
and spin parts $L_{ab}$ and $S_{ab}$ do not annihilate $\Psi$
separately. The difficulties are also illustrated by the 
following very rough argument. For superstring theory, the 
equation $Q\Psi =0$ is schematically of the form
$\left({\partial\over\partial X}+X\right)\psi =0$,
which has the square integrable solution $\psi= e^{-\br 12X^2}$,
whereas for the supermembrane it has the schematic form
$\left({\partial\over\partial X}+X^2\right)\psi =0$ which suggests
a non-normalizable solution of the type $\psi= e^{-\br 13X^3}$.
Of course, this is by no means a counterargument against the existence 
of a normalizable groundstate, precisely because the above equations 
do {\em not} factorize, and all the components are coupled. But the argument 
does indicate that the wave function must be very clever indeed if 
it is to evade this apparent obstruction to square integrability. 

Even if it is not possible to find the ground state wavefunction
explicitly, evidence for (or against) its existence can be supplied
by calculating the Witten index \cite{Yi:1997eg,Sethi:1997pa} 
(see also \cite{Smilga:1986jg} for earlier work in this direction). 
This index is formally defined as the difference between the number 
of bosonic and the number of fermionic states:
$$
I_W = \#(\hbox{bosonic states})-\#(\hbox{fermionic states})
$$
For a well behaved theory with a discrete spectrum every bosonic state
with nonzero energy can be transformed into a fermionic state by
acting on it with the supercharge and vice versa. So the
contribution to the index from the states of nonzero energy
vanishes. This reasoning does not hold for states with zero energy
since they are annihilated by the supercharge. So a non-vanishing
Witten index indicates the existence of zero energy states,
i.e. ground states. 

A regularized formula handy for calculations is
$$I_W = \lim_{\beta\to\infty}\Tr\left[(-1)^Fe^{-\beta H}\right].$$
Here, $F$ is the fermion number operator. The exponential factor makes
the trace well defined but does not disturb the result since it only
affects positive energies. In fact, for well behaved theories with a
discrete spectrum (where the Hamiltonian $H$ is a Fredholm operator)
the expression is independent of the ``inverse temperature'' $\beta$
and can be evaluated at small $\beta$ with the help of field theory
perturbative methods.

If, however, the spectrum is continuous and if there is no mass gap 
things are quite a bit more tricky: The above argument does not
imply that the bosonic and fermionic densities of states cancel for
positive energies. So we really have to evaluate the index for large
$\beta$, a region that is not accessible to perturbation
theory. Following \cite{Yi:1997eg,Sethi:1997pa}
we therefore write the index as
$$
I_W = \lim_{\beta\to 0}\Tr\left[(-1)^Fe^{-\beta H}\right]+\delta I
$$
where so-called defect is defined as 
$$
\eqalign{\delta I&=\int_0^\infty d\beta{d\over d\beta}\Tr
(-1)^Fe^{-\beta H}\cr
&= -\int_0^\infty d\beta\Tr(-1)^FHe^{-\beta H}\cr
&= -\int_0^\infty dE \left[\rho_B(E)-\rho_F(E)\right].\cr}
$$
with the bosonic and fermionic spectral densities $\rho_B$ and $\rho_F$,
respectively. The difference of these two spectral densities would
vanish for Fredholm operators. The defect is difficult to compute and 
few rigorous results are so far available.

With the help of the heat-kernel representation we write the 
bulk term as a finite dimensional integral: Defining 
$$
\eqalign{{\cal Z}_{D,N}&=\int\prod_{\mu,A}{dX_\mu^A\over\sqrt{2\pi}}
\prod_{\alpha,A}d\Psi_\alpha^A\, 
\exp\left\{\br 12\tr[X_\mu,X_\nu]^
2+\tr\Psi\Gamma^\mu[X_\mu,\Psi]\right\}\cr
&=\int\prod_{\mu,A}{dX_\mu^ A\over\sqrt{2\pi}}\, 
{\cal P}_{D,N}(X)\exp\left\{\br12\tr[X_\mu,X_\nu]^ 2\right\}\cr}
$$
we have
$$
I_W = {\cal F}_N^{-1}{\cal Z}_{D,N}+\delta I.
$$ 
where the normalization factor ${\cal F}_N$ is given by
\cite{Krauth:1998xh,Krauth:1998yu}\ 
$$
{\cal F}_N ={ {2^{{N(N+1)}\over 2} \pi^{{N-1}\over 2}}\over
              {2\sqrt{N} \prod_{k=1}^{N-1} k!} } 
$$
and ${\cal P}_{D,N}(X)$ is the Pfaffian 
$$
{\cal P}_{D,N}(X) =
\hbox{Pf}(-if_{ABC}\Gamma^{\mu}_{\alpha\beta}X_\mu^C).
$$
It is the square root of the determinant of an antisymmetric 
${\cal N}(N^2-1)\times{\cal N}(N^2-1)$ matrix and therefore 
a homogeneous and $SO(D)\times SU(N)$ invariant polynomial of 
degree $(D-2)(N^2-1)$ in the $X$'s, which is explicitly known 
only in a very limited number of cases. 

The analytic results for $SU(2)$ read:
$${\cal Z}_{D,2} = \sqrt{8\pi}\cases{0&for $D=3$\cr 1/4&for $D=4$\cr
1/4&for $D=6$\cr 5/4&for $D=10$\cr}$$
The normalization is ${\cal F}_2=\sqrt{8\pi}$. This suggests 
$$\delta I=-\br 14$$
which was indeed found in \cite{Sethi:1997pa}. This means that 
for $SU(2)$ the index is
$$I_W=\cases{0&for $D=4,6$\cr 1&for $D=10$\cr}.$$
implying the existence of a normalizable groundstate for the
maximally supersymmetric $SU(2)$ matrix model, but not for the
lower dimensional ones.

Instanton calculations suggest the following generalization for
$SU(N)$:\cite{Green:1997tn} 
%The above result suggests the following conjecture\cite{Green:1997tn}
$$
{\cal Z}_{D,N} = {\cal F}_N\cases{0&for $D=3$\cr 1/N^2&for
$D=4,6$\cr\sum_{m|N}{1\over m^ 2}&for $D=10$\cr}\gln\princpart
$$
and
$$
\delta I=\cases{0&for $D=3$\cr 
          -{1/ N^2}& for $D=4,6$ \cr
          -\sum_{m|N\atop m\ge 2}{1\over m^2}&for $D=10$\cr}\gln\defekt
$$
A proof of the first part of this conjecture was given very
recently in \cite{Moore:1998et}, while an independent test using 
Monte-Carlo integration has been performed in \cite{Krauth:1998xh}.
If the second part of the above conjecture could also be verified
this would indicate that the above result for $SU(2)$ generalizes
to any $N$, such that a normalizable ground state exist for all $N$
when $D=10$, but none for $D=4,6$. 

A curious feature of \princpart\ is the non-analytic nature
of the $N\to\infty$ limit in ten dimensions: Because for all $N$ 
$$
1<\sum_{m|N}{1\over m^2} < \sum_{m=1}^\infty{1\over
m^2}={\pi^2\over 6}
$$
we can find increasing subsequences $\{N_j\}$ of the natural numbers,
such that any given real number $1\le c\le\pi^ 2/ 6$ can be
obtained as a limiting value:
$$
\lim_{j\to\infty}\sum_{m|N_j}{1\over m^2}=c
$$
This shows that, even in the absence of large $N$ divergences, 
the $N\rightarrow\infty$ limit could be quite subtle.
The non-analytic behavior may also be a special feature of the 
supersymmetric theory and $D=10$, whereas there are indications 
\cite{Staudacher} that the bosonic matrix model exhibits a more 
regular behavior\footnote{The non-supersymmetric bosonic model 
(obtained simply by dropping the Pfaffian from the above integral) 
was recently investigated in \cite{Krauth:1998yu}, where numerical 
evidence was presented that the integrals exist if 
$$
N>\frac{D}{D-2}.
$$}. Incidentally, the validity of the formulas \princpart\ and 
\defekt\ indicates that one should be able to analytically 
solve the so-called IKKT model:\cite{Ishibashi:1996xs}
$$
\eqalign{
{\cal Z}_{IKKT}(\beta) &= \sum_{N=1}^\infty \ignore{C_N}
\sum_{m|N}{1\over m^2}e^{-\beta N}  \cr
&= \sum_{m,n=1}^\infty {1\over m^2}e^{-\beta m n} 
 = \sum_{m=1}^\infty {1\over m^2} \frac{1}{e^{m\beta}-1}   \cr} 
$$
Apart from the fact that analyticity has been restored, the
main interest of this model resides in the fact that it treats
all $N$ simultaneously in a kind a grand canonical partition
function, albeit only for the completely reduced model. It would 
be remarkable if one could extend these considerations to the
full (1+0)-dimensional matrix model in terms of a chemical
potential for the D0 particles.

Besides the work on Witten indices, there is now a growing body
of rigorous results, although so far mostly for certain truncations. 
For the $SU(2)$ model of \cite{deWit:1988ig}, a rigorous proof of 
non-existence is now available \cite{Froehlich:1997df}; for further
extensions of these results, see \cite{Hoppe:1997wc}. In \cite{Porrati:1997ej}
the question of normalizable states was investigated by means of
a deformed version of the equations defining the supersymmetric 
groundstate (where the deformation in particular introduces 
``mass terms'' modifying the continuous spectrum to a discrete one). 
Yet another approach based on an investigation of the asymptotic 
nature of the groundstate for $SU(2)$ was initiated
in \cite{Halpern:1997fv}, and further pursued in \cite{Hoppe:1997fr}.

However, even if we were able to establish the existence of massless 
states for arbitrary $N$ we would not yet be done: after all, we would 
also like to know {\em what they look like}! If M(atrix) theory is really 
the fundamental theory it is claimed to be, knowledge of its groundstate
wave function would profoundly alter our outlook on the world. 
\medskip

\noindent{\bf Acknowledgements:} We would like to thank M. Staudacher
and B. de Wit for helpful comments on the manuscript.

%
% Hier war das Literaturverzeichnis
%
%\bibliography{membran}

\end{document}
\bye